\documentclass[pdflatex,iicol,sn-mathphys-num]{sn-jnl}          % twocolumn

\usepackage[utf8]{inputenc}
\usepackage{amsmath, amssymb, mathtools}
\usepackage{braket}
\usepackage{upgreek}
\usepackage{graphicx} 
\usepackage{dcolumn}% Align table columns on decimal point
\usepackage{bm}% bold math
\usepackage{xcolor}
\usepackage{booktabs} % \toprule / \midrule / \bottomrule
\usepackage{numprint} % to truncate + round digits in table
\usepackage{lmodern} % more font sizes
\usepackage{csquotes}
\usepackage{breakurl}

\usepackage{subfigure}

\usepackage{hyperref} % must be last
\hypersetup{
    pdftitle={Accurate theoretical methods for photoionization of H2 molecules},
}

\newcommand{\mbf}[1]{\mathbf{#1}}
\newcommand{\mrm}[1]{\mathrm{#1}}
\newcommand{\op}[1]{\hat{\mathrm{#1}}}
\newcommand{\labeq}[1]{\label{eq:#1}}
\renewcommand{\refeq}[1]{Eq.\,(\ref{eq:#1})}
\newcommand{\labsec}[1]{\label{sec:#1}}
\newcommand{\refsec}[1]{Sec.\,\ref{sec:#1}}
\newcommand{\labfig}[1]{\label{fig:#1}}
\newcommand{\reffig}[1]{Fig.\,\ref{fig:#1}}
\newcommand{\labtbl}[1]{\label{tbl:#1}}
\newcommand{\reftbl}[1]{Table\,\ref{tbl:#1}}
\newcommand{\rme}{\mathrm{e}}
\newcommand{\rmd}{\mathrm{d}}
\newcommand{\im}{\mathrm{i}}

\gdef\orcidlogo{\includegraphics[width=3mm]{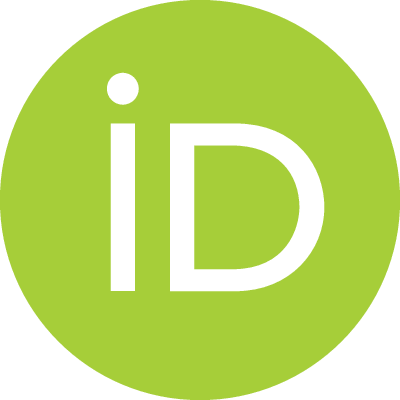}}

\begin{document}

\title{Accurate theoretical methods for photoionization of \texorpdfstring{$\mrm{H}_2$}{H2} molecules}
\author*[1]{\fnm{Hakon} \sur{Volkmann} \orcid{0009-0005-4758-3551}}
\affil[1]{\orgdiv{AG Moderne Optik}, \orgname{Institut f\"ur Physik, Humboldt-Universit\"at zu Berlin}, \orgaddress{\street{Newtonstr.~15}, \postcode{12489} \city{Berlin} \country{Germany}}}
\affil[2]{\orgdiv{Institute for Nuclear Physics}, \orgname{Westf\"{a}lische Wilhelms-Universit\"{a}t M\"{u}nster}, \orgaddress{\street{Wilhelm-Klemm-Str.~9}, \postcode{48149} \city{M\"{u}nster}, \country{Germany}}}
\email{hakon.volkmann@physik.hu-berlin.de}
\author[1, 2]{\fnm{Jannis} \sur{Sch{\"u}rmann} \orcid{0009-0005-3890-2653}}
\email{jannis.schuermann@physik.hu-berlin.de}
\author[1]{\fnm{Alejandro} \sur{Saenz} \orcid{0009-0002-8886-2336}}
\email{alejandro.saenz@physik.hu-berlin.de}

\date{\today}

\abstract{Single-photon ionization cross sections of molecular hydrogen in the electric dipole limit are numerically computed by employing three independent methods.
Both time-dependent and -independent approaches within the clamped-nuclei approximation at equilibrium internuclear distance are used, featuring the explicit time-propagation of the time-dependent Schr{\"o}dinger equation and newly developed multi-channel configuration-interaction free-boundary as well as complex-scaling methods using an explicitly correlated geminal basis set.
The found results show convincing mutual agreement despite their entirely different fundamental formulations.
They further highlight the challenges in bringing together different theoretical predictions from literature with the experimental data at high photon energies.
Overall, the novel CI-based approach demonstrates fast and controllable convergence while being able to provide full channel-resolved information, indicating the need for more accurate experimental data.}

\maketitle

%\SetWatermarkText{DRAFT v8}
%\SetWatermarkScale{1.0}
%\SetWatermarkLightness{0.95}

\section{Introduction} 

Photoionization of hydrogen molecules belongs to the group of supposedly well-understood processes that were subject of a large number of works in the past. 
The simplicity of this molecular system allows for thorough theoretical investigations that highlight the specific influence of electronic correlation~\cite{sct:wait17}, nuclear motion~\cite{sfm:foer14} and non-adiabatic couplings~\cite{sfm:pala07}, and multi-center interference effects~\cite{sct:cohe66, sct:nagy04} on molecular light-matter dynamics, just to name a few.

While still in the infancy of quantum mechanics, it was Bethe who recognized~\cite{sct:beth30} that ionization cross sections of fast electron impact, for which the Born approximation is applicable, leads (in the limit of small scattering angles) to a qualitative behavior remarkably similar to that of single-photon ionization.
As a consequence, ``quasi'' photoionization cross sections (PICS) could be determined ``by proxy'' from electron-scattering experiments~\cite{sct:wiel72} that yield larger cross section signals, which was essential at times when tunable light sources still had been very scarce.
Concomitantly, it is, of course, also possible to gain information on small-angle electron-impact ionization cross sections by considering those of photoionization. 

In recent years, within the context of the KATRIN neutrino-mass experiment~\cite{nu:katr19,nu:katr25}, a renewed interest arose in highly accurate electron-scattering cross-section data on T$_2$ molecules.
In order to accurately determine the kinetic energy of an electron emitted during $\upbeta^-$ decay, inelastic electron-scattering events off T$_2$ molecules prior to detection have to be taken into account~\cite{sct:aker21}.
It has been found that there is a satisfactory agreement between the cross section of photoexcitation and low-energy photoionization with that of electron impact.
As the KATRIN experiment demands for particularly accurate input data from theory, a thorough analysis of the different computational approaches is due.

Several approaches for treating this problem have been devised over the past years.
The aim of this work is to re-visit the photoionization problem of H$_2$ molecules at equilibrium internuclear distance $R=1.4\ a_0$, for which both a reasonably large amount of theoretical reference data is available~\cite{sct:flan65,sct:khar68,sct:mart74,sct:ford75,sct:flan77,sct:onei78,sct:lucc81,sct:itik83,sct:rase83,sct:rase84,sct:coll84,sct:rich84,sct:rich86,sct:hara86,sct:tenn86,sct:cace93,sct:sanc97,bsp:mart99,sct:liu04,sct:fojo04,sct:nagy04,sct:borb07,sct:sanz07,sct:toff16} and experimental verification~\cite{sct:back76,sct:gall88,sct:koss89,sct:sams94,sct:lati95,sct:glas07} is possible.
The applicability of this problem in the limit of small-angle electron scattering ultimately serves as a validation of methods employed for the proper electron-scattering calculations that will be published in a separate work.
Accordingly, a comparison of photoionization cross sections resulting from three different methods is presented.
Two of these approaches, namely the free-boundary (FB) and the complex-scaling geminals (CSG) method, are based on the time-independent formalism.
For the former approach, a configuration interaction (CI) basis of ionic orbitals has been used, while explicitly correlated basis functions are employed for the latter.
The third approach is based on a perturbative description, utilizing the solution of the explicitly time-\emph{dependent} Schr\"odinger equation (TDSE) instead.
However, an ionic CI representation closely related to the one used for the FB method is employed for the numerical representation of the time-dependent wavefunction.
As a verification of self-consistency, the symmetry-resolved first-order (Thomas-Reiche-Kuhn) as well as higher-order sum rules are given.
First, these methods are briefly showcased in Secs. \ref{sec:fbm} and \ref{sec:gcsm}, respectively.
Furthermore, the time-dependent description is considered in \refsec{tda}, before further details on the numerical treatment are provided in \refsec{numer}.
The sum rules and photoionization cross sections obtained from all three approaches as well as the photo-excitation cross sections from the latter two methods are presented in \refsec{results}, before a discussion is given in \refsec{discussion}, finally followed by a summary in \refsec{summary}.

%%%%%%%%%%%%%%%%%%%%%%%%%%%%%%%%%%%%%%%%%%%%%%%%%%%%%%%%%%%%%%

\section{Theoretical methods}
\labsec{theory}

In this section, a brief overview of the different theoretical foundations underlying the applied numerical methods are given.
As this work intents to be of comparative nature, due discussions of the details of all of the methods are not provided here.
However, as most of the methods are well-established in the literature, further information on their respective implementation details can be found in the provided references.
Note, if not stated otherwise, the atomic unit system with $\hbar=m_e=4\pi\epsilon_0=1$ is being adopted.

\subsection{Free-boundary method}
\labsec{fbm}

The direct approach for obtaining PICS data requires a set of channel-resolved scattering states of (molecular) symmetry $\Gamma$.
Such states have to be computed on a sufficiently dense energy grid for each channel $a$ open at energy $E$ and will be denoted $\Psi^{(\Gamma)}_{aE}$ in the following.
Hence, $\Psi^{(\Gamma)}_{aE}$ is an (improper) eigenstate of the molecular Hamiltonian
\begin{equation}
 \labeq{ham}
   \op H \;=\; \op H_\mrm{ion} \,+\, \frac{1}{2}\op{p}^2_0 + \op V(\mbf r_0, \mbf r)\,,
\end{equation}
where, in the context of this work, the nonrelativistic ionic Hamiltonian $\op H_\mrm{ion}$ denotes the (single-particle) Hamiltonian of the H$_2{}^+$ molecular ion (clamped at inter-nuclear distance $R$) and $\op V(\mbf r_0, \mbf r)$ contains its interaction with the ejected photoelectron at position $\mbf r_0$ and momentum $\mbf p_0$.
In particular, $\op V$ contains both, the Coulomb attraction between the photoelectron and the nuclei as well as the repulsion from the bound electron at position $\mbf r$.

The FB method discretizes the (non-breakup) position-space wavefunction $\Psi^{(\Gamma)}_{aE}$ in terms of a symmetry-adapted close-coupling (CC) basis set.
It combines the single-particle basis functions $\{B_i(\mbf r)\}$, vanishing at the edge $\partial{\mathcal D}$ of the single-particle computational domain $\mathcal D$ such that
\begin{equation}
   \labeq{zero_bc}
   B_i\big|_{\partial\mathcal D} \;=\; 0\qquad\forall\ i\,,
\end{equation}
with suitable single-particle basis functions $\chi_j$ for the photoelectron coordinate $\mbf r_0$ of the same boundary condition
\begin{equation}
   \chi_j\big|_{\partial\mathcal D} \;=\; 0\qquad\forall\ j\,.
\end{equation}
The CC expansion of $\Psi^{(\Gamma)}_{aE}$ then reads
\begin{align}
 \begin{split}
  \labeq{cc}
  \Psi^{(\Gamma)}_{aE}(\mbf r_0, \mbf r) \,=\, \sum\limits_{ij} &c_{ij}^{(a)}\, \op{\mathcal A}^{(\Gamma)}\left[\chi_i(\mbf r_0)\, B_j(\mbf r)\right]\\
  &+\, f_a(\mbf r_0)\phi_a(\mbf r)\,,
 \end{split}
\end{align}
%If solved numerically, this task involves some variant of a discretization procedure of the corresponding time-independent Schr\"odinger equation, \emph{e.g.} by employing either collocation/grid or Galerkin (\emph{i.e.} ``basis set expansion'') methods.
%In terms of large electronic structure computations, particularly the latter approach poses a common and well-understood method in order to obtain both bound states and a set of scattering states for a range of energies compatible with the spatial density of discretization and the chosen computational box-size.
%There, typically a basis is chosen that conveniently features homogeneous (or zero-)boundary conditions on the (potentially asymptotic) box edge, agreeing with the typical asymptotic behavior of those bound states fitting into the box.
%The set of energies of the obtained scattering states corresponds to those solutions (or linear combinations thereof) that coincidentally fulfill the zero-boundary conditions at the box edge.
%As such, control over the precise value of energy a scattering solution is sought for can only be indirectly achieved by tweaking the boundary condition, \emph{e.g.} by performing a box-size variation.
with the unknown, energy-dependent coefficient vector $\mbf c^{(a)}$ and the quasi-projection $\op{\mathcal A}^{(\Gamma)}$ onto the irreducible representation $\Gamma$  (up to a normalization). 
The single-particle function $f_a$ is chosen such that it is the only function in the basis that takes a finite value at $\partial\mathcal D$ while $\phi_a$ is an ionic eigenstate (of energy $\epsilon_a$) of $\op H_\mrm{ion}$ that characterizes the inelastic scattering channel $a$.
The computational domain $\mathcal D$ is chosen sufficiently large such that it is reasonable to approximate $\phi_a|_{\partial\mathcal D}=0$, rendering the exchange term at $\partial\mathcal D$ negligible.
Furthermore, $f_a$ has to be chosen such that its product with $\phi_a$ is compatible with the total symmetry $\Gamma$.
%Instead of enforcing the boundary condition to be zero on the box edge and subsequently losing control over the energy at which the solutions are obtained, the free boundary methods rather enforce a certain value energy $E$ at which scattering solutions for all channels can be obtained.

The homogeneous scattering problem 
\begin{subequations}
 \begin{equation}
  \left(\op H - E\right)\,\Psi^{(\Gamma)}_{aE} \;\equiv\; \op A(E)\,\Psi^{(\Gamma)}_{aE} \;=\; 0
 \end{equation}
 with inhomogeneous boundary condition
 \begin{equation}
  \Psi^{(\Gamma)}_{aE}(\mbf r_0, \mbf r)\Big|_{\mbf r_0\in \partial\mathcal D} \;=\; \phi_a(\mbf r)\,f_a(\mbf r_0)
 \end{equation}
\end{subequations}
and freely chosen values of $E>\epsilon_0$ (with $\epsilon_0$ being the lowest-lying threshold) can be cast into an equivalent inhomogeneous problem
\begin{subequations}
 \begin{equation}
  \op A(E)\, \Psi^{(\Gamma)'}_{aE} \;=\; -\op A(E)\, \phi_a\,f_a\,,
 \end{equation}
 with homogeneous boundary condition
 \begin{equation}
   \Psi^{(\Gamma)'}_{aE}\Big|_{\mbf r_0\in\partial\mathcal D} \;=\; 0\,,
 \end{equation}
\end{subequations}
whose respective solutions are related by
\begin{equation}
  \Psi^{(\Gamma)}_{aE} \;=\; \Psi^{(\Gamma)'}_{aE} \,+\, \phi_a\,f_a\,.
\end{equation}
Employing the finite expansion \refeq{cc} and imposing the Galerkin condition on the residual yields the linear problem
\begin{equation}
    (\mbf H\, -\, E\,\mbf S)\,\mbf c^{(a)} \;=\; \mbf q^{(a)}(E)\,.
\end{equation}
Here, $\mbf H_{ij,kl}=\Braket{\chi_iB_j|\op H\, \op{\mathcal A}^{(\Gamma)}|\chi_kB_l}$ and $\mbf S_{ij,kl}=\Braket{\chi_iB_j|\op{\mathcal A}^{(\Gamma)}|\chi_kB_l}$ denote the Hamiltonian and overlap matrices while
\begin{equation}
  q^{(a)}_{ij}(E) \;=\; \Braket{\chi_iB_j|\op{\mathcal A}^{(\Gamma)}\left( E-\op H \right)|f_a\phi_a}
\end{equation}
is one of $N_\mrm{chan}$ freely, but linear-independently chosen boundary vectors.
The scattering wavefunction obtained this way is an (arbitrary) linear combination of all open channel solutions $\mbf c^{(a)}$, which subsequentially has to be transformed into the $S$-matrix basis in order to recover physical cross sections.
By fitting all channel solutions to the asymptotic standing-wave behavior near $\partial\mathcal D$~\cite{sct:bart96},
 \begin{equation}
  \begin{split}
  \labeq{fit}
  &\Psi^{(\Gamma)}_{aE}(\mbf r_0, \mbf r) \;\underset{r_0\to\infty}{\sim}\;
  \op{\mathcal A}^{(\Gamma)}\, \sum_{b}\, \frac{1}{r_0}\sqrt{\frac{1}{\pi\, k_b}}\, \phi_b(\mbf r)\, \times\\
  &\times\; Y^{m}_{\ell_b}(\op{\mbf r}_0)\, \left[ A_{ab}\, F_{\ell_b}(k_b r_0) \,+ \, B_{ab}\, G_{\ell_b}(k_b r_0) \right]\,,
  \end{split}
\end{equation}
the matrices $\mbf A$ and $\mbf B$ are recovered.
In \refeq{fit}, the single-particle (ir-)regular Coulomb wavefunctions~\cite{sct:gord28, sct:volk26} are denoted as $F_{\ell}$ ($G_{\ell}$), whereas $Y^{m}_{\ell}$ refers to the the quasi-angular channel and $k_b=\sqrt{2(E-E_b)}$ to channel momentum with threshold energy $E_b$.
After applying $\mbf A^{-1}$, one obtains the Heitler reactance (or $K$) matrix solution as $\mbf K = \mbf A^{-1} \mbf B$, while the $S$-matrix solution may be obtained from the transformation
\begin{align}
 \begin{split}
    \tilde{\Psi}^{(\Gamma)}_{aE} &= \sum_b \left[\left(\mbf 1+\im\mbf K\right)^{-1}\,\mbf A^{-1}\right]_{ab} \Psi^{(\Gamma)}_{bE} \\
    &= \sum_b \left[(\mbf A+\im\mbf B)^{-1}\right]_{ab} \Psi^{(\Gamma)}_{bE}\,.
 \end{split}
\end{align}
%The same is true for the fixed boundary solutions; there however, this linear transformation is hampered by the fact that only a single solution instead of a set of a channel solutions is given at a certain value of energy, required to perform the linear transform.
In order for the procedure to be sufficiently accurate, the FB approach requires a box-size chosen such that $\Psi^{(\Gamma)}_{aE}\Big|_{\partial\mathcal D}$ is reasonably close to the asymptotic behavior given by \refeq{fit}.
That is, all inter-channel couplings are neglected beyond $\partial\mathcal D$, as is also typically assumed in $R$-matrix approaches~\cite{sct:tenn10}.
However, in contrast to the $R$-matrix method, no matching conditions involving logarithmic derivatives are required for the FB approach.

Evidently, the boundary condition is chosen ``freely''~\cite{bsp:bros92a, bsp:bros92b, sct:brag92, sct:lamb98, bsp:bach01, sfa:niko01a, bsp:hart02, sct:niko06} in the sense that the approximated wavefunction is not forced to vanish at the boundary, as it is typically the case for $L^2$ methods adopting a basis with zero-boundary conditions \footnote{The term ``free'' might not be entirely fitting in this situation, despite the formal similarities to the previous methods carrying the same name.
As a matter of fact, the numerical solution is rather sought with respect to a fixed, \emph{inhomogeneous} boundary condition that typically leads to a numerical solution in terms of a linear combination over all channels.
The term ``free'' hence rather refers to the freely-chosen value of energy than the boundary condition itself.}.
Both the least-squares and the FB approaches are viable options to obtain the scattering solutions.
While the boundary condition of the former method can be considered to be truly free, as it is self-consistently determined during a variational least-squares minimization of the residual, the latter method directly enforces the (non-zero) inhomogeneous boundary condition at the box edge (as seen above) in terms of an $L^2$ Galerkin problem, ultimately leading to a set of $N_\mrm{chan}$ linear problems.
The FB approach was chosen here for its ease of integration with the already existing, fixed-boundary code~\cite{bsp:vann04} at hand.
A more detailed discussion of this method is beyond the scope of this work and hence will be subject of future works.

Once the channel-resolved and $S$-matrix-normalized scattering states $\tilde{\Psi}_{aE}$ are determined, the orientation-averaged (single-)photon PICS is determined by~\cite{sct:star23}
\begin{align}
 \begin{split}
  \labeq{dip_cross}
  &\frac{\rmd\sigma^\mrm{(L/V)}}{\rmd E}\\
  &=\; \frac{4\pi^2}{3c}\sum_{p \in\{x,y,z\}}\sum\limits_{\Gamma,a} \bigg|\Braket{\tilde{\Psi}^{(\Gamma)}_{aE}|\op{\mu}_p^\mrm{(L/V)}|\Psi_g}\bigg|^2\,,
 \end{split}
\end{align}
with $\op\mu_p^\mrm{(L)} = \sqrt{E}\, \mbf e_p\cdot\hat{\mbf r}$ being the length-form and $\op\mu_p^\mrm{(V)} = \mbf e_p\cdot\nabla/\sqrt{E}$ the velocity-form dipole-coupling operators, $\mbf e_p$ the light-polarization direction as well as $\Psi_g$ the ground-state of $^1\Sigma_\mrm{g}^+$ symmetry.
In this work, the intermolecular axis is chosen to coincide with $\hat{\mbf e}_z$, hence perpendicularly polarized light $\mbf e_p \perp \op{\mbf e}_z$ selects $\Gamma=\phantom{\,}^1\Pi_\mrm{u}$ states, while light polarized as $\mbf e_p \,||\, \hat{\mbf e}_z$ selects scattering states $\Psi^{(\Gamma)}_{aE}$ of $\Gamma=\phantom{\,}^1\Sigma_\mrm{u}^+$ symmetry.
For an isotropic gas of diatomic molecules one has
\begin{align}
 \begin{split}
 \labeq{pics}
 \frac{\rmd\sigma^\mrm{(L,V)}}{\rmd E} \;=\; \frac{4\pi^2}{3c} \left( D^\mrm{(L/V)}_{||}(E) + 2\,D^\mrm{(L/V)}_\perp(E) \right)
 \end{split}
\end{align}
with
\begin{equation}
 \labeq{ampl_fbm}
  D^\mrm{(L/V)}_{p}(E) \,=\, \sum_a \bigg|\Braket{\tilde{\Psi}^{(^1\Sigma_\mrm{u}^+\,/\, ^1\Pi_\mrm{u})}_{aE}|\op{\mu}_p^\mrm{(L/V)}|\Psi_g}\bigg|^2\,.
\end{equation}

\subsection{Time-dependent approach}
\labsec{tda}

An alternative possibility for obtaining the PICS, distinct from the one laid out in Sec.~\ref{sec:fbm}, is to employ a time-dependent method~\cite{sct:sanz07,sct:awas05,bsp:dumi07,sct:apal02}. 
Instead of computing the time-independent molecular scattering states directly, it is possible to consider a time-dependent process, where a probe laser pulse of reasonably narrow spectral width centered around $E$ ionizes a single electron.
For sufficiently low intensities, the overall single-photon ionization process can be treated perturbatively, yielding the corresponding cross section up to lowest-order perturbation theory (LOPT), together with \refeq{pics}, as
\begin{align}
 \begin{split}
 \labeq{ampl_tdse}
    D&^\mrm{(L/V)}_{p}(\omega)\\
    &=\; Y^\mrm{(L/V)}_{p}(\omega)\,\times\ \left(\ \int_{-T/2}^{T/2} \frac{I(t)}{\omega}\ \mathrm dt\ \right)^{-1},
 \end{split}
\end{align}
where $Y^\mrm{(L/V)}_{p}(\omega)$ denotes the total ionization yield for polarizations $p\in\{\,||,\,\perp\}$, $I(t)$ the time-dependent laser intensity profile, $\omega = E$ the central pulse energy, and $T$ the probe-pulse duration.
The ionization yield is calculated by subtracting the sum of the populations for all bound (ground and excited) states from the total probability according to
\begin{align}
 \begin{split}
  \labeq{yield}
    Y&^\mrm{(L/V)}_{p}(\omega)\\
    &= \; 1\; - \sum\limits_{\{\text{bound } n\}} \big|\,d^\mrm{(L/V)}_{p,n}(\omega, t=T/2)\,\big|^2\,.
 \end{split}
\end{align}
The time-dependent amplitudes $d^\mrm{(L/V)}_{p,n}(\omega,t)$ result from a discrete expansion of the full time-dependent two-electron state
\begin{align}
  \begin{split}
  \Psi_{p,\omega}^\mrm{(L/V)}(&\mathbf r_1, \mathbf r_2, t; R)\\
  &=\; \sum_{n} d^\mrm{(L/V)}_{p,n}(\omega,t)\, \psi_n^{(p)}(\mathbf r_1, \mathbf r_2; R)
 \end{split}
\end{align}
in terms of a time-independent basis of field-free eigenstates $\psi_n^{(p)}$ of \refeq{ham} with energies $E_n$.
The final state $\Psi(\mbf r_1, \mbf r_2, t=T/2; R)$ is computed for each value of $E$ numerically by solving the truncated discretized versions of the TDSE
\begin{equation}
  \im\partial_t\Psi_{p,\omega}^\mrm{(V)} \;=\; \left[\op H \,-\, \mbf A_{p}(\omega,t)\cdot\op{\mbf p}\right]\, \Psi^\mrm{(V)}_{p,\omega}
\end{equation}
for the velocity gauge and
\begin{equation}
  \im\partial_t\Psi_{p,\omega}^\mrm{(L)} \;=\; \left[\op H \,+\, \mbf E_{p}(\omega,t)\cdot\op{\mbf r} \right]\, \Psi^\mrm{(L)}_{p,\omega}
\end{equation}
for the length gauge, where the molecule, initially prepared in the ground-state of $\op H$, interacts with a laser pulse in a gauge obeying $\mbf E_p(\omega,t) = -\partial_t \mbf A_p(\omega,t)$ with
\begin{equation}
  \mbf A_{p}(\omega,t) = \frac{A_0}{\omega}\cos^2\left(\frac{\pi t}{T}\right)\sin(\omega t)\,\mbf{e}_{p}
\end{equation}
for parallel and perpendicular polarization vectors $\mbf e_p$.

An essential advantage of this approach over the FB method lies in the implicit inclusion of the channel structure for \refeq{ampl_tdse}, in contrast to the explicit summation required in \refeq{ampl_fbm}.
Furthermore, its treatment of excitation and ionization channels is based on a common footing, as the excitation yield can be obtained in a straight-forward way similar to \refeq{yield}.
Within the context of the Born-Oppenheimer approximation, this is particularly convenient in the ionization threshold region, where this approach yields a smoothly joined spectrum, whereas for the other methods, the results need to be combined with excited-state cross-section data and then \enquote{sewed} together manually.
Another handy property of this approach is the possibility of directly imposing or emulating a finite experimental resolution on the spectral width of the probing pulse.
In contrast, the data obtained from the time-independent methods have to be artificially smeared, \emph{e.\,g.}, via Gaussian convolution, in order to resemble experimental results more closely.
Further details about the method employed in this work can be found in previous works~\cite{sct:awas05,bsp:dumi07,sct:apal02}, where the entire procedure has been successfully applied.

\subsection{Geminal complex-scaling method}
\labsec{gcsm}

A third option to obtain the total transition amplitudes is to apply the complex-scaling method, see, \emph{e.\,g.}, Refs.~\cite{csm:agui71, csm:bals71, csm:nutt69, csm:resc75,  csm:rein82, csm:froe84a, nu:froe93}.
The action of the originally unitary real-scaling transformation on wavefunctions
\begin{equation}
   \left[\op U(\zeta)\psi\right](\mathbf r_1, \mathbf r_2)=\mathrm e^{-3\zeta}\psi\left(\mathrm e^{\zeta}\, \mathbf r_1, \mathrm e^{\zeta}\, \mathbf r_2\right)\,,
\end{equation}
is analytically continued into the complex plane by allowing the dilation $\zeta$ to adopt complex values $\zeta=\im\theta$ with $\theta\in\mathbb R$ being the dilation angle.
It has been shown that the poles of the transformed resolvent
\begin{align}
 \begin{split}
  \labeq{cs_resolv}
  \op G^{(\theta)}(E) \;&\equiv\; \left[E - \op U(-\im\theta)\, \op H\, \op U(\im\theta)\right]^{-1} \\
   &=\; \left(E \,-\, \op H^{(\theta)}\right)^{-1} 
 \end{split}
\end{align}
remain independent of $\theta$ within a certain range of values~\cite{csm:agui71,csm:bals71}, while rotating the continuum branch cuts around their respective thresholds into the complex plane by an angle of $-2\theta$.
This property allows for a direct numerical evaluation of the discretized resolvent at real-valued continuum energies, for which the branch cuts are transformed into a series of complex-valued poles, conveniently avoiding the singularities. 
In particular, it is possible to recover the optical oscillator strength (OOS) density without requiring an explicit numerical representation of the scattering states \cite{sct:saen93, sct:saen96, dia:saen03},
\begin{align}
  \begin{split}
  \labeq{csm_oos}
  &\frac{\rmd f}{\rmd E} = \frac{2E}{\pi}\ \mathrm{Im}\, \Braket{\psi_0^{(-\theta)}|\op z^{(\theta)}\, \op G^{(\theta)}(E)\, \op z^{(\theta)}|\psi_0^{(\theta)}}
  \end{split}
\end{align}
by employing a well-known variant of the Kramers-Kronig dispersion relations.

In this work, the explicitly correlated basis proposed by Ko{\l}os and Wolniewicz~\cite{dia:kolo68} is used to compute dipole transition spectra for a field orientated parallel to the molecular ($\op z$) axis.
A code originally written by Pachucki \emph{et al.}~\cite{dia:pach09,dia:pach16}\ has been extended such that the Hamiltonian matrix elements $H_{ij}$ are evaluated as functions of real-valued $\zeta$ on a certain grid of values.
In order to obtain the complex-scaled matrix elements $H^{(\theta)}_{ij}$, low-order interpolation polynomials over the $\theta$ grid are computed and subsequently analytically continued to complex values $\zeta=\im\theta$.
After fully solving the system $\mathbf H^{(\theta)}\mbf c=\epsilon^{(\theta)}\mathbf S\mathbf c$ (with overlap matrix $\mathbf S$), a discrete spectral representation
\begin{equation}
  \op G^{(\theta)}(E) \;=\; \sum_{j=1}^K \frac{\Ket{\psi_j^{(\theta)}}\Bra{\psi_j^{(-\theta)}}}{E_j^{(\theta)}-E_0^{(\theta)}-E}
\end{equation}
of \refeq{cs_resolv} with discretized eigenstates $\Ket{\psi_j^{(\theta)}}$ and \emph{complex} energies $E_j^{(\theta)}$ can be obtained and inserted into \refeq{csm_oos}, together with a proper numerical representation of the complex-scaled ground state $\Ket{\psi_0^{(\pm\theta)}}$.
Note that not only the Hamiltonian needs to be analytically continued into the complex plane, but the dipole-operator matrix-elements as well.

\section{Numerical details}
\labsec{numer}
All approaches presented in \refsec{theory} have in common that they have been implemented within the prolate spheroidal coordinate system $(\xi,\eta,\varphi)$.
As the H$_2{}^+$ wavefunctions are separable in these coordinates, numerical methods face no challenge in modelling the electron-nuclear cusps at the Coulomb singularities, even for large internuclear separations.
Furthermore, all methods employed in this work are in some sense of Galerkin type and hence operate in terms of a finite basis.
Thus, a quick rundown of the two kinds of basis functions, namely ionic orbital CI and explicitly correlated geminals, are presented in the following.

\subsection{Configuration interaction}
\labsec{ci}
Despite the fundamentally different approaches, both the TDSE and the FB methods in this work still rely on a CI basis in terms of a symmetry-adapted combination of ionic orbitals,
\begin{equation}
  \labeq{ci}
  \Phi_{ij}(\mbf r_1, \mbf r_2) \;=\; \op{\mathcal A}^{(\Gamma)}\left[ \phi_i(\mbf r_1)\,\phi_j(\mbf r_2)\right]\,,
\end{equation}
with $\phi_i$ solving the single-electron time-independent Schr\"odinger equation
\begin{equation}
  \op H_\mrm{ion}\,\phi_i(\mbf r) \;=\; E_i\,\phi_i(\mbf r)\,.
\end{equation}
The ionic orbitals $\phi_i$ themselves are expressed in terms of a separable product basis of $N_\xi-1=200$ uniformly distributed B splines $b_i$ of order 7 for the $\xi$ coordinate and $N_\eta = 10$ of order 5 for the $\eta$ coordinate.
Together with the function $\exp(\im m\varphi)$ for the azimuth angle $\varphi$, they are given by
\begin{equation}
  \phi_i(\xi,\eta,\varphi) \;=\; \frac{1}{\sqrt{2\pi}}\,X_i(\xi)\,Y_i(\eta) \,\rme^{\im m\varphi}\,,
\end{equation}
with
\begin{equation}
  X_i(\xi) \;=\; (\xi^2-1)^{|m|/2}\,\sum_{j=1}^{N_\xi-1} x_{ij}\, b^{(\xi)}_j(\xi)
\end{equation}
for the coordinate $\xi$ and
\begin{align}
 \begin{split}
  Y_i(\eta) \;=\; (1&-\eta^2)^{|m|/2}\, \sum_{j=1}^{N_\eta} y_{ij}\, \times\\
  &\times\, \left[b^{(\eta)}_j(\eta) + (-1)^{|m|+\wp}\, b^{(\eta)}_{N_{\eta}-j+1}(\eta)\right]
 \end{split}
\end{align}
for $\eta$ with expansion coefficients $x_{ij}, y_{ij}$ and parity quantum number $\wp = 0$ for \emph{gerade} and $\wp = 1$ for \emph{ungerade}  states as well as the magnetic quantum number $m$. 
Further details on the employed CI implementation can be found in Ref.~\cite{bsp:vann04}.

As the B splines are chosen to fulfill zero-boundary conditions, the CI basis-functions used in \refeq{ci} inherit the same property.
The ramifications of this circumstance are different in the two methods employed.
For the TDSE, this effectively amounts to the introduction of infinitely high potential walls around the molecule.
As such, it is vital for them to lie sufficiently far away from the origin such that the invariably occurring reflection artefacts of the emitted electron wavefunction are negligible.
In turn, for the FB method, only the summation term in the r.h.s. of \refeq{cc} can be expressed in terms of CI basis-functions \refeq{ci}.
For the boundary function $f_a$, the choice
\begin{equation}
  f_a(\xi,\eta,\varphi) \;=\; \frac{(\xi^2-1)^{|m_a|/2}}{\sqrt{2\pi}}\,b^{(\xi)}_{N_\xi}(\xi)\, Y_a(\eta)\, \rme^{\im m_a\varphi}
\end{equation}
with a single B spline $b^{(\xi)}_{N_\xi}$ that is non-vanishing at $\xi_\mrm{max}=100$ $a_0$ has been made.
The linear independence of this boundary function is therefore a direct consequence of the orthogonality of $\phi_a$ and $Y_a$.

The structure of the CI series included in the basis evidently depends on the kind of state one wishes to describe.
The tension inherent to \refeq{dip_cross} between providing a basis flexible enough that it can simultaneously describe both locally confined, correlation-sensitive ground states as well as asymptotically far-reaching and highly oscillatory scattering states is a primary reason for the choice of the B-spline CI approach.
While the ground-state CI contains a large set of low-energy state combinations, the continuum states typically demand for \emph{complete} series of ionic states for each scattering channel $\phi_a$ to be included in the expansion.
Furthermore, for each of the two parities (\emph{gerade} and \emph{ungerade}) and lowest pseudo-angular channels, series of the $8\times 8$ energetically lowest-lying bound states are included in the basis in order to approximate the doubly-excited bound-state character of the continuum wavefunction for energies in the vicinity of Fano resonances.
The $N_\mrm{chan}$ channel functions $\phi_a$ are chosen to be the energetically lowest states for the given value of internuclear separation at $R=1.4$ $a_0$, where the 6 ionic channels 1s$\sigma_g$, 2p$\sigma_u$, $\pm$2p$\pi_u$, 2s$\sigma_g$, and 3p$\sigma_u$ below the 6th threshold are considered for the FB method computations.
Furthermore, for each of these 6 ionic channels, the lowest 5 pseudo-angular channels (with 0, 2, \ldots, 8 for even and 1, 3, \ldots, 9 nodes for odd $\eta$ wavefunctions $Y_i(\eta)$, respectively) have been included.
All of the overall 30 channels then feature a full series of $N_\xi-1=200$ ionic orbitals.
Finally, in order to properly describe the non-zero boundary condition, the remaining highest 5 pseudo-angular channels of each ionic channel also contain series for the 5 energetically \emph{highest} ionic orbitals, resulting in a total amount of 5582 configurations for the $^1\Sigma^+_\mrm{u}$ and 6672 for the $^1\Pi_\mrm{u}$ symmetries, respectively. 
Further increasing the number of channels have not shown to significantly alter the results, which also corroborates the observations made in Ref.~\cite{sct:fojo04}.
The corresponding configuration series used for the ground-state have been found heuristically and are explicitly given in \reftbl{gsci} of appendix \ref{sec:ci_gs}.

\subsection{TDSE method}

Compared to the FB method, the requirements on the ionic orbital basis used for the TDSE approach are more demanding, as the wavepacket of the ionized electron has to be described within a finite box for a sufficiently long pulse duration.
Since both the FB and TDSE methods share a common electronic structure description, they can be characterized by the same numerical parameters that got introduced in \refsec{ci}.
As such, a box size of $\xi_\mrm{max}=300$ $a_0$ is covered by $N_\xi-1=300$ B splines of order 8 in $\xi$ direction and $N_\eta=10$ B splines of order 6 in $\eta$ direction.
The configuration series employed in the solution of the TDSE, as for the FB method, has to include those series that represent the most-relevant channels as well as doubly-excited configurations for the Fano resonances.
Their precise definition is given in \reftbl{tdse_ci_spu} in appendix \ref{sec:ci_gs}.

\subsection{Ko{\l}os-Wolniewicz basis functions (geminals)}
% kolos wolniewicz basis functions
The Ko{\l}os-Wolniewicz basis functions~\cite{dia:kolo64}
\begin{align}
 \begin{split}
  \Phi&_{\mathcal N}(\mbf r_1, \mbf r_2)\\
  &=\; r_{12}^{n_0}\eta_{1}^{n_1}\eta_{2}^{n_2}\xi_{1}^{n_3}\xi_{2}^{n_4}\, \rme^{-\alpha \xi_1 -\bar{\alpha} \xi_2 - \beta \eta_1 - \bar{\beta} \eta_2}
  \labeq{eq:kolos_wolniewicz_basis_function}
 \end{split}
\end{align}
are known to produce very accurate results for the ground and first
excited states of two-electron diatomic molecules~\cite{dia:kolo65,sct:woln95,sct:stas02,sct:woln03}, due to the
direct incorporation of the inter-electronic distance $r_{12}$.
% definition of a base
Here, a single basis function $\Phi_{\mathcal N} $ can be identified by an exponent
quintuple $\mathcal N = (n_0,n_1,n_2,n_3,n_4)$ and a set of non-linear parameters
$\{\alpha, \bar{\alpha}, \beta, \bar{\beta}\}$.
The H$_2$ wavefunction $\Psi$ can be represented as a linear combination
of Ko{\l}os-Wolniewicz basis functions with coefficients $c_\mathcal{N}$ by enforcing the proper symmetry
\cite{dia:pach16}
\begin{equation}
  \Psi = \frac{1}{4}\left(1 \pm I_{12}\op{P}_{12}\right)\left(1 \pm I_\mrm{AB}\op{P}_\mrm{AB}\right) \sum_{\mathcal N} c_{\mathcal N}\, \Phi_{\mathcal N}
\end{equation}
with respect to the permutation of the two electrons 1 and 2 (through operator
$\op{P}_{12}$ with eigenvalue I$_{12}$) and the permutation of the two nuclei A and B
(through operator $\op{P}_\mrm{AB}$ with eigenvalue I$_\mrm{AB}$). As already
mentioned, for the computation, an own extension of the H2SOLV code
\cite{dia:pach16} was used.

In order to increase both readability and reproducibility, the notion of a so-called \emph{base} as introduced in Ref.~\cite{sct:schn24} is adopted here.
A base is characterized by the set of non-linear parameters $\{\alpha, \bar{\alpha}, \beta, \bar{\beta}\}$ (which are kept constant for all basis functions in said base)
and a set of unique exponent quintuples $\mathcal N = (n_0,n_1,n_2,n_3,n_4)$.
Furthermore, this set of exponent quintuples can be uniquely identified by the rank $\Omega$ of the base, for which each quintuple fulfills $\Omega \geq \sum_i n_i$.
This allows to reduce the complete definition of a base to
$\{\alpha, \bar{\alpha}, \beta, \bar{\beta}; \Omega\}$.
Departing from Ref.~\cite{nu:froe93}, a multi-base complex-scaling approach was implemented (similar to the dual-base approach in Ref.~\cite{dia:brum26}), \emph{i.\,e.},~the use of more than one base with different non-linear parameters, combined into a single eigenvalue problem.
In order to overcome numerical near-linear dependencies inherent to the finite-precision representation of matrix elements involving Ko{\l}os-Wolniewicz basis functions, variable precision linear algebra within the H2SOLV code has been utilized.
Furthermore, instead of using the von-Neumann expansion as in Ref.~\cite{nu:froe93}, the H2SOLV code employs a Taylor expansion in $R$ for the evaluation of the required integrals.

% method / complex scaling parameters
Apart from these modifications, the complex-scaling approach is implemented exactly in the same way as originally introduced in Ref.~\cite{nu:froe93} for the Ko{\l}os-Wolniewicz basis functions.
Hence, the same dilation parameters have been used for the interpolation of matrix elements, for which the number of points $N_{\rho} = 7$ is varied linearly between $\rme^\zeta =$ 0.97 and 1.03 and a polynomial of degree $N_{p} = 4$ was used for the extrapolation.
For all complex-scaling results shown in this work, the dual-base
%\begin{align}
%  \begin{split}
%   %&\qquad \{\alpha,\, \bar{\alpha},\, \beta,\, \bar{\beta},\, \Omega\} \in \\
%    &\big(\{0.218,\, 0.558,\, 0.094,\, 0.24;\, 6\},\ \{ 0.1,\, 0.5,\, 0.1,\, 0.0;\, 6 \} \big)
%  \end{split}
%\end{align}
\begin{align}
  \begin{split}
   \{\alpha,\, \bar{\alpha},\, \beta,\, \bar{\beta},\, \Omega\} \,\in\, \big(\{&0.218,\, 0.558,\, 0.094,\, 0.24;\, 6\},\\
   \{ &0.1,\, 0.5,\, 0.1,\, 0.0;\, 6 \} \big)
  \end{split}
\end{align}
has been used. 
Here the tuple notation (comma-separated elements, enclosed by parentheses)
has been used in order to concatenate the two single bases into a dual-base definition.
The first non-linear parameters were obtained by dividing the T$_2$
parameters from Table I in Ref.~\cite{nu:froe93} by the internuclear separation
$R = 1.4\,a_0$, effectively transforming them into the $R$-independent non-linear
parameter convention used in the H2SOLV code~\cite{dia:pach16}.
The second set was acquired from a grid search for a base with more diffuse
$\xi_1$ and $\xi_2$ exponential parameters, \emph{i.e.},~where $\alpha$ and $\bar{\alpha}$ are smaller.
From said grid, the base with the highest density of states in the
energy interval from 0.0 to 0.3 a.u.~was chosen.
This is the energy range which includes
the series of higher-lying resonances.

% scaling angles theta
The complex-scaling angle $\theta$ was varied in steps of 0.01 radian between 0
and 0.78, verifying sufficient stability against the rotation into the complex
plane. However, for the complex-scaling spectrum shown in this work, only values
between 0.18 and 0.4 radian were used.
For each energy, the evaluation of \refeq{csm_oos} has been performed for
the most-stable value of the scaling angle $\theta$.
An overall trend towards higher values of $\theta$ at energies close to the
ionization threshold and decreasing values for higher energies has been observed.

% evaluation of the results
\npdecimalsign{.} % use '.' as the decimal sign
\nprounddigits{3} % Round of all numbers in the table after the third digit
\tabcolsep2mm     % add some padding around the columns

\begin{table*}[t]
    \centering
    \caption{
      Comparison of the discretized (bound state) sum rules obtained for the
      B-spline CI and for the geminals at $R = 1.4\,a_0$. Also shown are the values
      taken from Ref. \cite{sct:woln95,sct:stas02,sct:woln03}
      and Ref.~\cite{sct:yan98}.
    }
    \begin{tabular}{ln{1}{3}n{1}{3}n{1}{3}n{1}{3}n{1}{3}n{1}{3}n{1}{3}n{1}{3}}
        \toprule
        & {CI$^\mrm{V}$} & & & {Gem$^\mrm{L}$} & {\cite{sct:woln95,sct:stas02,sct:woln03}} &  &  & \cite{sct:yan98}\\
        \midrule
        S$_k$    &
        ${^1\Sigma_\mrm{u}^+}$ &
        ${^1\Pi_\mrm{u}}$    & 
        isotropic       & 
        ${^1\Sigma_\mrm{u}^+}$ &
        ${^1\Sigma_\mrm{u}^+}$ &
        ${^1\Pi_\mrm{u}}$    &
        isotropic      &
        isotropic \\
        \midrule
        % CI-374610 == Free-Boundary CI scaled to TDSE box (rmax210)
        %        S$_{-2}$ & 1.6994606496 & 1.020063298  & 3.7395872456 & 1.6954909020 & 1.658286 & 0.948477 & 3.555240 & 3.784 \\
        %        S$_{-1}$ & 0.8212653226 & 0.511543916  & 1.8443531546 & 0.8240529916 & 0.802114 & 0.474340 & 1.750794 & 1.884 \\
        %        S$_{0}$  & 0.3993005876 & 0.2578595813 & 0.9150197503 & 0.4027869823 & 0.389829 & 0.238013 & 0.865855 & 0.943 \\
        %        S$_{1}$  & 0.195460702  & 0.130710154  & 0.4568810100 & 0.1981238283 & 0.190462 & 0.119861 & 0.430183 & 0.474 \\
        %        S$_{2}$  & 0.0963954133 & 0.06665485   & 0.2297051133 & 0.098133559  & 0.093599 & 0.060595 & 0.214789 & 0.239 \\
        % CI-374616 == Free-Boundary CI scaled to TDSE box + 1-100 : 1-100 terms (rmax210)
        S$_{-2}$ & 1.7069714943 & 1.0300337546 & 3.7670390036 & 1.6954909020 & 1.658286 & 0.948477 & 3.555240 & 3.784 \\
        S$_{-1}$ & 0.8243614176 & 0.5163561306 & 1.857073679  & 0.8240529916 & 0.802114 & 0.474340 & 1.750794 & 1.884 \\
        S$_{0}$  & 0.400550618  & 0.2601938553 & 0.9209383286 & 0.4027869823 & 0.389829 & 0.238013 & 0.865855 & 0.943 \\
        S$_{1}$  & 0.195950616  & 0.1318489753 & 0.4596485666 & 0.1981238283 & 0.190462 & 0.119861 & 0.430183 & 0.474 \\
        S$_{2}$  & 0.0965788696 & 0.0672140726 & 0.2310070149 & 0.098133559  & 0.093599 & 0.060595 & 0.214789 & 0.239 \\
        \bottomrule
    \end{tabular}
    \labtbl{sum_rules_bound}
\end{table*}

\begin{table*}
    \centering
    \caption{
      Symmetry-resolved continuum oscillator-strength sum rules for B spline
      based CI, TDSE, and CSG. Comparison of combined isotropic oscillator-strength
      sum rules for B-spline based CI and TDSE to isotropic continuum
      sum rules from literature, see text for more details.
    }
    \begin{tabular}{ln{1}{3}n{1}{3}n{1}{3}n{1}{3}n{1}{3}n{1}{3}n{1}{3}n{1}{3}n{1}{3}n{1}{3}n{1}{3}}
        \toprule
        & {CI$^\mrm{V}$} & & {TDSE$^\mrm{V}$} & & {CSG$^\mrm{L}$} & {CI$^\mrm{V}$} & {TDSE$^\mrm{V}$} & {\cite{sct:yan98,sct:liu04}} & {\refeq{corrected_yan01},\cite{sct:yan98}} \\
        \toprule
        &
        ${^1\Sigma_\mrm{u}^+}$ &
        ${^1\Pi_\mrm{u}}$ &
        ${^1\Sigma_\mrm{u}^+}$ &
        ${^1\Pi_\mrm{u}}$ &
        ${^1\Sigma_\mrm{u}^+}$ & isotropic & isotropic & isotropic & isotropic \\
        \midrule
        S$_{-2}$ & 0.4380330783 & 0.513272372  & 0.4381578802908428 & 0.5133298444453483 & 0.434670408 & 1.4645778223 & 1.4648175691815395 & 1.466 & 1.4498259854855422 \\
        S$_{-1}$ & 0.3316251736 & 0.427912684  & 0.3310156786982555 & 0.4267325332708239 & 0.328499159 & 1.1874505416 & 1.1844807452399033 & 1.182 & 1.171797644701554  \\
        S$_{0}$  & 0.2678455546 & 0.4028230806 & 0.2637035365489168 & 0.3960863036648211 & 0.265120230 & 1.073491716  & 1.055876143878559  & 1.063 & 1.0565631141899914 \\
        S$_{1}$  & 0.2643499383 & 0.4817233946 & 0.2275635086737288 & 0.4340859836455938 & 0.260632979 & 1.2277967276 & 1.0957354759649167 & 1.219 & 1.214606606043044  \\
        S$_{2}$  & 0.757400572  & 1.1923136060 & 0.2238159993123399 & 0.5976305171728477 & 0.575214856 & 3.142027784  & 1.4190770336580352 & 3.582 & 3.579397058170677  \\
        \bottomrule
    \end{tabular}
    \labtbl{sum_rules_continuum}
\end{table*}

\subsection{Sum rules}

For verification purposes, it is useful to compute the OOS between the initial ground and the \emph{bound} final states as
\begin{subequations}
\begin{align}
 \begin{split}
 \labeq{oosl}
  f&^{(p,\,\mrm L)}_n\\
  &=\; \frac{2(E_n-E_g)}{3}\, \Bigg|\Braket{\psi_n^{(p)}|\mbf e_p\cdot\sum_{j=1}^2 \op{\mbf r}_j|\Psi_g}\Bigg|^2
 \end{split}
\end{align}
in the length form as well as
\begin{align}
 \begin{split}
 \labeq{oosv}
  f&^{(p,\,\mrm V)}_n\\
  &=\; \frac{2}{3(E_n-E_g)}\, \Bigg|\Braket{\psi_n^{(p)}|\mbf e_p\cdot\sum_{j=1}^2 \nabla_j|\Psi_g}\Bigg|^2
 \end{split}
\end{align}
\end{subequations}
in the velocity form with ground-state energy $E_g$.
The molecular eigenstates $\psi_n^{(p)}$ (of energy $E_n$) are of the same polarization-dependent dipole-allowed symmetry as laid out in the end of \refsec{fbm} and are obtained from a diagonalization of the finite-basis CI-/CSG-representation of the molecular Hamiltonian, of which only a finite number $n$ from the infinite amount of Rydberg states is taken into account.
Instead, if scattering states are employed, the discrete, $n$-indexed quantities in Eqs.~(\ref{eq:oosl}) and (\ref{eq:oosv}) turn into OOS densities $\rmd f^{(p,\, \mrm{L/V})}/\rmd E$, \emph{cf.} \refeq{csm_oos}, w.\,r.\,t.~continuous values of energy $E$.
Both forms are equivalent in the limit of a complete basis, but may differ appreciably, if approximations for the ground or final states are utilized.
As special values of OOS moments
\begin{align}
 \begin{split}
 \labeq{sum_rules}
     S^{(p)}_k = &\sum_{\{\text{bound } n\}} (E_n-E_0)^k\,f^{(p)}_n \\
     &+\, \int_0^\infty (I+E)^k\,\frac{\rmd f^{(p)}}{\rmd E}\,\rmd E
 \end{split}
\end{align}
for $k\in\{-2,-1,0,1,2\}$ and ionization potential $I$ are known to fulfill exact sum rules~\cite{sct:yan98}, they allow to assess the quality, self-consistency, and completeness of the (truncated) computational basis.
By these means, the summation \refeq{sum_rules} can be conveniently extended to run over the discretized continuum (or unbound) states as well, as the normalization of the latter contains the weights of the implicit quadrature rule for the spectral measure~\cite{sct:rein79}, which is well-known from, \emph{e.\,g.}, Stieltjes-imaging techniques.

%%%%%%%%%%%%%%%%%%%%%%%%%%%%%%%%%%%%%%%%%%%%%%%%%%%%%%%%%%%%%%%%%%%%%%%%%%%%%%%%%%%%%%%

\section{Results}
\labsec{results}

\subsection{Dipole oscillator sum rules}

\begin{table*}[ht]
    \centering
    \caption{
      Comparison of length and velocity forms used for the isotropic total
      discretized sum rules obtained within the B spline based CI. For comparison,
      the results from Ref.~\cite{sct:woln93} are given as well.
    }
    \begin{tabular}{ln{1}{3}n{1}{3}n{1}{3}n{1}{3}n{1}{3}n{1}{3}n{1}{3}}
        \toprule
        & {CI$^\mrm{L}$} & & & {CI$^\mrm{V}$} & & & {\cite{sct:woln93}} \\
        \toprule
        S$_k$    &
        ${^1\Sigma_\mrm{u}}$ &
        ${^1\Pi_\mrm{u}}$ &
        {iso. tot.} &
        ${^1\Sigma_\mrm{u}}$ &
        ${^1\Pi_\mrm{u}}$ &
        {iso. tot.} &
        {iso. tot.} \\
        \midrule
        S$_{-2}$ & 2.1377556423 & 1.543459242  & 5.2246741263 & 2.1374937283 & 1.5333356699 & 5.2041650683 & 5.180 \\
        S$_{-1}$ & 1.1529909196 & 0.9467365586 & 3.046464037  & 1.1528904963 & 0.9394566006 & 3.0318036976 & 3.036 \\
        S$_{0}$  & 0.6689404636 & 0.6702657193 & 2.0094719023 & 0.6671461423 & 0.660682662  & 1.9885114663 & 2     \\
        S$_{1}$  & 0.4729653276 & 0.6429462873 & 1.7588579023 & 0.4598106403 & 0.6124335486 & 1.6846777376 & 1.701 \\
        S$_{2}$  & 0.9401851739 & 1.429880282  & 3.799945738  & 0.8537959853 & 1.258968456  & 3.3717328973 & 3.851 \\
        \bottomrule
    \end{tabular}
    \labtbl{sum_rules_gauge}
\end{table*}

In the following, the orientation-averaged oscillator strength moment $S_k^{(p)}$, as defined in \refeq{sum_rules}, will be denoted as \emph{isotropic}, whereas the combination of bound and continuum contributions are being called \emph{total}.
Note, the superscript $^{(p)}$ indicating the polarization direction is dropped for the oscillator strength moment $S_k^{(p)}$ throughout this section, as it is irrelevant for isotropic sum rules or redundant, if the final-state molecular symmetry ($\Sigma$ for $||$, $\Pi$ for $\perp$) is provided.
Furthermore, all sum rules shown are marked with superscript $^\mrm{L}$ for the length and $^\mrm{V}$ for the velocity forms.
Columns labelled as CI$^\mrm{L/V}$ were obtained with the TDSE-CI, i.e the CI used during the TDSE propagation (which is structured similar to the FB-CI).
However, due to the larger box requirements of the TDSE method, a larger number of B splines (compared to the FB method) is required in order to obtain satisfactory results.
This further limits the completeness of the TDSE-CI.
The sum rules computed with CI and CSG stem from a discretized summation,
while the TDSE sum rules result from a spline-interpolation based integration.

% results bound state table
The bound-states sum rules are presented in \reftbl{sum_rules_bound}.
The data taken from Ref.~\cite{sct:woln95,sct:stas02,sct:woln03} contain only the six lowest ${}^1\Sigma_\mrm{u}^+$ and four lowest ${}^1\Pi_\mrm{u}$ bound states, \emph{i.e.},~they do not include the
infinite series of Rydberg states below the ionization threshold.
Note, the geminal and the CI sum rules include a box-discretized representation of these Rydberg states.
In spite of using velocity (CI) and length forms (CSG), both methods yield 
matching results and appear to extend the sum rules from
Ref.~\cite{sct:woln95,sct:stas02,sct:woln03} converged beyond the agreement to Ref.~\cite{sct:yan98}.

% what / where ?
In \reftbl{sum_rules_continuum}, a comparison of the various methods adopted
for the continuum contribution to the sum rules is shown.
% what can be seen
An excellent agreement is found between the methods for the energy momenta $S_k$,
$k \in \{-2,-1,0\}$. This is furthermore substantiated by the observation that
CSG is formulated in length form, while the shown CI-based sum rules are given in velocity form.
For $k = 1$, CSG and CI still agree very well, while the TDSE starts to deviate.
For $k = 2$, all three methods differ significantly from each other.
Note, higher energy momenta $S_k$ are increasingly sensitive to the high-energy part
of the photoionization spectrum.
The deviation between the CI-based methods and CSG can be explained by the following two arguments:
First, both bases used for the CSG are optimised for an energy region rather close to the ionization threshold. 
Second, the FB-CI focuses only on the first 200 eV of the spectrum. The deviations of the TDSE results,
especially when compared to the CI, can be explained by the fact that the TDSE
spectrum was only obtained up to 4 a.u., since only states up to 5 a.u.~entered
the time-propagation. As is discussed below in the context of \reffig{Su_log}, this
apparently does not affect the spectrum significantly, but might be responsible
for the deviations starting at approximately 75 eV.

% Isotropic sum rules
The isotropic continuum sum-rules shown in the lower half of
\reftbl{sum_rules_continuum} demonstrate the same pattern of agreement to the literature,
as the upper half shows within itself. That is, very good agreement is found for the lower
momenta $S_k$, $k \in \{-2,-1,0\}$, whereas the agreement is worse for $k=1$ (especially for the TDSE) and
a significant departure from the other methods can be seen for $k=2$.
The better agreement between the $k=0$ TDSE sum-rules and those from literature (compared to the CI method) is likely rooted in the finite spectral width of the pulse.
To a certain extent, the resulting smearing is artificially (and conveniently) mimicking the effect of nuclear-motion broadening, in spite of being a purely single-$R$, that is, fixed-nuclei result.

The effects of choosing the length or velocity forms for the sum rules obtained from the CI and TDSE
methods is further summarized in \reftbl{sum_rules_gauge} and validated by the sum rules given in Ref.~\cite{sct:woln93}.
% what can be seen, results gauge comparison
As the comparison of length and velocity forms is based on the same CI, they involve exactly the same wavefunctions upon evaluation.
Thus, what is distinguishing the forms from each other is merely the way in which the (approximate) wavefunctions ``probe'' the two different interaction operators.
Again, for the lower energy momenta $S_k$, $k \in \{-2,-1,0\}$, the different
forms show good agreement and deviate only on the sub-percent level from the
reference values given in Ref.~\cite{sct:woln93}. The agreement of the $^1\Sigma_\mrm{u}^+$
sum rules between CI and TDSE is one order of magnitude better than for the
$^1\Pi_\mrm{u}$ symmetry.
Again, this is probably rooted in the worse representation of the high-energy spectrum used for the TDSE compared to the CI method, as the contribution from the $^1\Pi_\mrm{u}$ symmetry
dominates in this case.
For $k=1$, both forms start to deviate clearly from the
reference values given in Ref.~\cite{sct:woln93}.
The difference is significant at $k=2$, but, as for the continuum sum-rules, can
be explained by the present focus on the low-energy part of the photoionization spectrum in the basis selection.

% Liu & Shemansky comparison
Concerning the continuum comparison between the employed methods, the use of the values from Table 6 of \cite{sct:yan98} as-is is not possible.
As Liu and Shemansky noted in Ref.~\cite{sct:liu04}, the H$_2$
sum rules given in Table 6 of Ref.~\cite{sct:yan98} for the 15.4 to 18 eV region (obtained from the erratum in Ref.~\cite{sct:yan01}) appear to be erroneous. 
They give an updated analytical expression, which was used here together with the remainder of Table 6 of Ref.~\cite{sct:yan98} in order to obtain the sum-rule results shown in column 9 of \reftbl{sum_rules_continuum}.
Note that the exact sum-rule values vary slightly in comparison to Ref.~\cite{sct:liu04}, as the integration was performed here based on a spline interpolation.

% Corrected Yan98 / Yan01
As an alternative, a corrected
form of the analytical expression given in the erratum of Ref.~\cite{sct:yan01}
\begin{align}
 \begin{split}
  \labeq{corrected_yan01}
  \sigma_{\mrm{H}_2}(E) \;=\; \Big(&1.191 - 197.448\, E^{-0.5} \,+\\
  &+\, 438.823\, E^{-1} \,-\, 260.481\, E^{-1.5}\,+\\
  &+\, 17.915\, E^{-2}\Big)\, 10^7\, {\rm barns} 
 \end{split}
\end{align}
was used.
It can be found by enforcing a cross section of 0 at 15.4 eV for Eq.\,(1) in Ref.~\cite{sct:yan01}. Note that this results in a discontinuity at 18 eV, albeit a rather small one.
The partial sum rules obtained with \refeq{corrected_yan01} for momenta of order $k = \{-2,-1,0,1,2 \}$ are
$S_k = 0.435,\, 0.271,\, 0.169,\, 0.105,\, 0.066,$ respectively.
Combining them again with the other sum-rule values from Table 6 of Ref.~\cite{sct:yan98} allows
for finding the corrected continuum sum-rule values shown in column 10 of \reftbl{sum_rules_continuum}.

\begin{figure*}[t]
  \subfigure[]{
  \labfig{su}
  \includegraphics[width=0.49\textwidth]{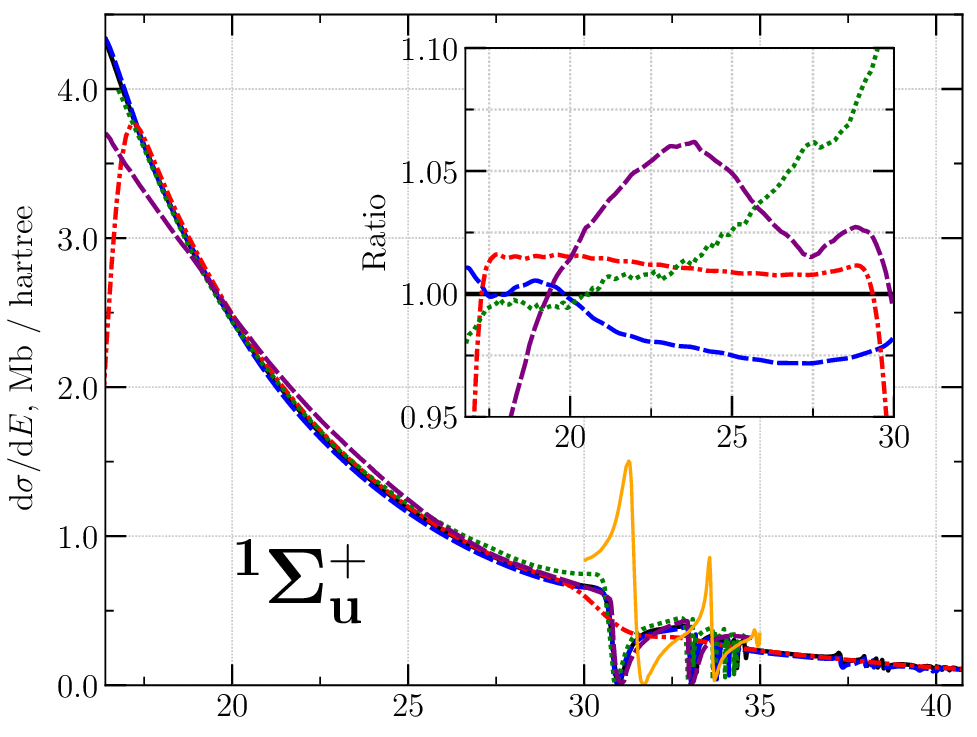}}
  \subfigure[]{
  \labfig{pu}
  \includegraphics[width=0.49\textwidth]{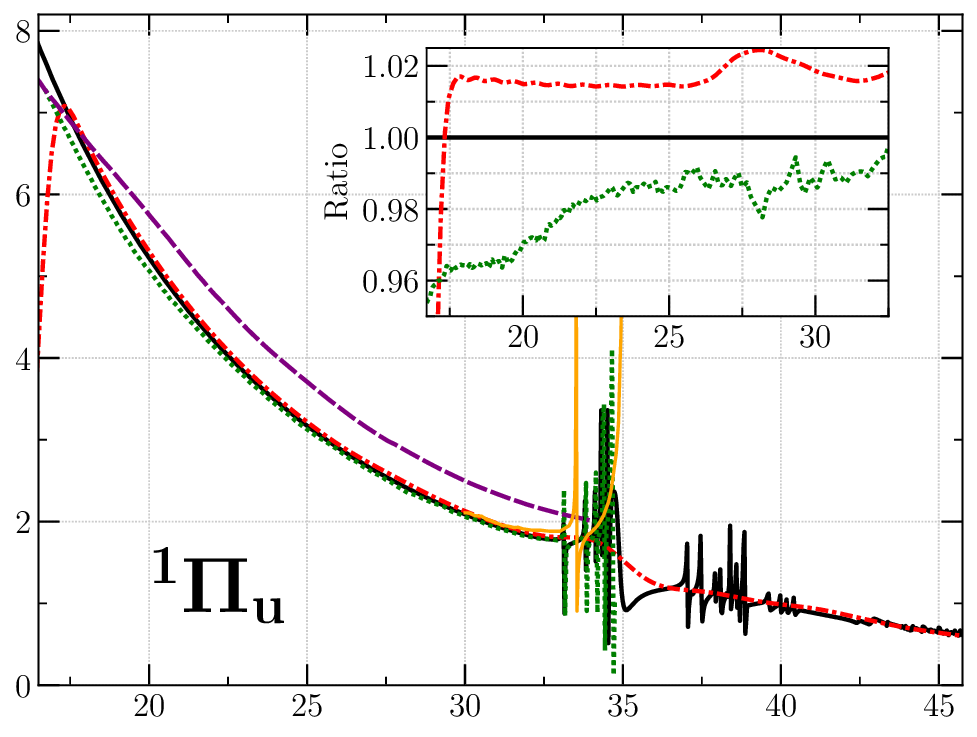}}
  \subfigure[]{
  \labfig{suz}
  \includegraphics[width=0.49\textwidth]{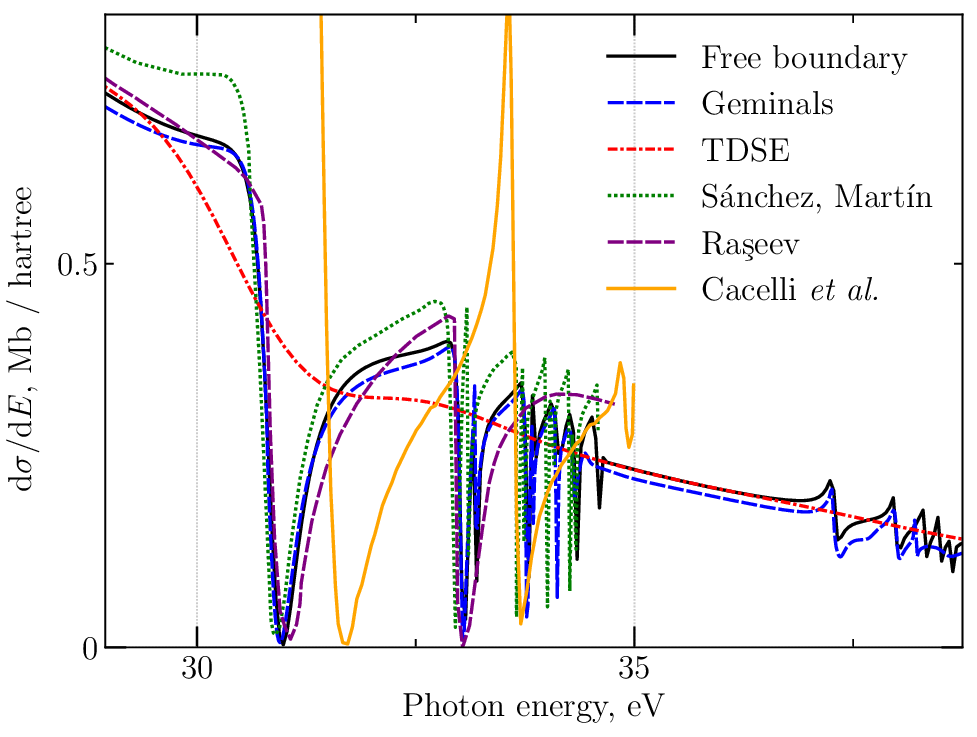}}
  \subfigure[]{
  \labfig{puz}
  \includegraphics[width=0.49\textwidth]{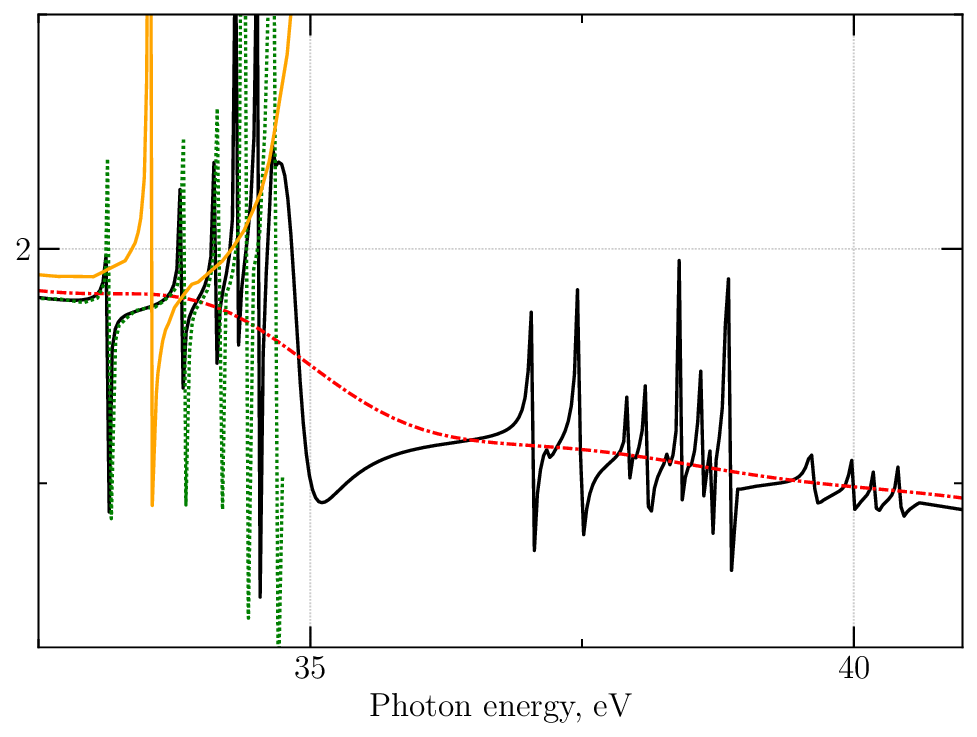}}
  \caption{Photoionization cross sections for parallel (left column, \subref{fig:su}, \subref{fig:suz}) and perpendicular (right column, \subref{fig:pu}, \subref{fig:puz}) oriented H$_2$ molecules, relative to the polarization direction.
  The lower figures in \subref{fig:suz}, \subref{fig:puz} depict an enlarged plot in the resonance region, the insets show the respective ratios relative to the free-boundary result.
  Solid (black) lines: FB-CI; dashed (blue): CSG; dash-dotted (red): TDSE; dotted (green): reproduced from S{\'a}nchez and Mart{\'i}n~\cite{sct:sanc97}; dashed (purple): reproduced from Ra\c{s}eev~\cite{sct:rase84}; solid (orange): reproduced from Cacelli \emph{et al}.~\cite{sct:cace93}.
  All curves shown are shifted in energy in order to match with the ionization potential of the FB result.
  Note that, due to the method's inherent finite resolution, the TDSE results are a spectral average over an energy window much broader than 
  most of the resonance widths.}
  \labfig{Su_v_Pu}
\end{figure*}

\subsection{Photoionization cross sections}
\labsec{pics}

Figure~\ref{fig:Su_v_Pu} further provides a general idea of the accuracy that can be achieved from the different theoretically obtained PICS for H$_2$ molecules.
For the most part of the considered energy range, relative deviations below 10\% can be observed both, for parallel (\reffig{su}) and for perpendicularly oriented molecules (\reffig{pu}).
This is particularly true for the detailed resonance line shapes more closely shown in Figs.~\ref{fig:suz} and \ref{fig:puz}.
In contrast to atoms, these are typically not directly observable in experiments due to the super-imposed smearing effect stemming from the vibrational degrees of freedom.
Nevertheless, they provide a useful benchmark for comparing the capabilities of the theoretical approaches with respect to describing electronic correlation of ground and doubly excited states.
The FB and CSG methods are in overall close agreement to one another, particularly in terms of resonance shapes and positions.
It has to be emphasized though that the explicitly included electron correlation in the CSG method allows for computing highly accurate ground-state energies at approximately $-1.1745$ a.u., to be compared to the FB ground-state energy of about $-1.1722$ a.u and the TDSE ground-state energy of approximately $-1.168$ a.u. (resulting from the larger box-size requirements of the time propagation).
Therefore, all results shown here have been shifted to the FB-CI scale for better comparability.
The TDSE method typically lies closest to the FB results, which may be not too surprising given that it can be regarded as a consequence of their common CI description.
It differs, however, more visibly in the resonance region, as the energy resolution in the TDSE method is more severely limited and not able to follow the intricate interference details, if not pulses of very long duration are considered.
The results of Sanch\'ez and Mart\'in~\cite{sct:sanc97} are also included, who used an $L^2$ Lippmann-Schwinger method based on a spherical-coordinate Hartree-Fock-orbital CI formulation.
It shows somewhat larger deviations near the resonances with respect to the present FB-CI and CSG results, but is otherwise found to be in good agreement.
The older results obtained in Refs.~\cite{sct:rase84} and \cite{sct:cace93} are included as well, which however show more severe deviations from the more recent works.
This is particular the case for the resonant line-shapes, for which a tendency of convergence over the years (accompanied by increasing computational power and complexity of the models) can be seen. 
Still, as the (ionic-orbital CI-based) FB method, the CSG approach, and the Hartree-Fock CI-based calculations in Ref.~\cite{sct:sanc97} all treat the electronic correlation within the doubly excited states in a different manner, a certain amount of remaining disagreement in the line shapes is not entirely surprising.

\begin{figure}[t]
  \includegraphics[width=1.0\columnwidth]{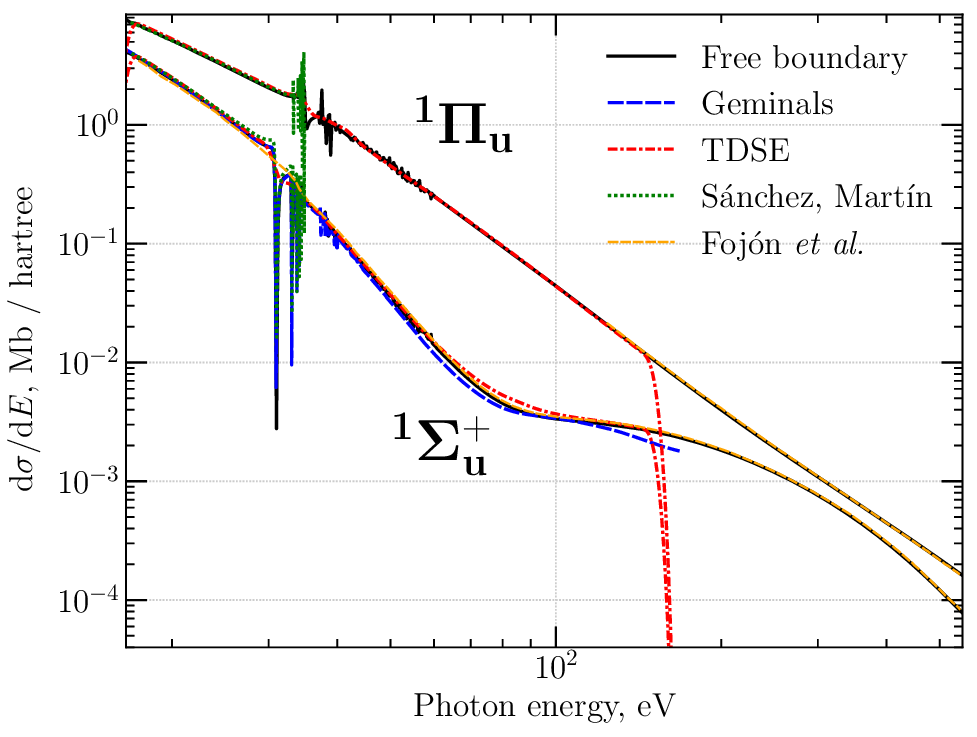}
  \caption{As \reffig{Su_v_Pu}, but on an extended (logarithmic) energy scale.
  The lines are specified as in the caption of \reffig{Su_v_Pu}.
  The upper lines correspond to a perpendicular, the lower to the parallel-aligned molecular axis, as labelled in the figure.
  Additionally, the result of Foj{\'o}n \emph{et al}.~\cite{sct:fojo04} is shown as dashed orange line.}
  \labfig{Su_log}
\end{figure} 

\begin{figure}
  \includegraphics[width=1.0\columnwidth]{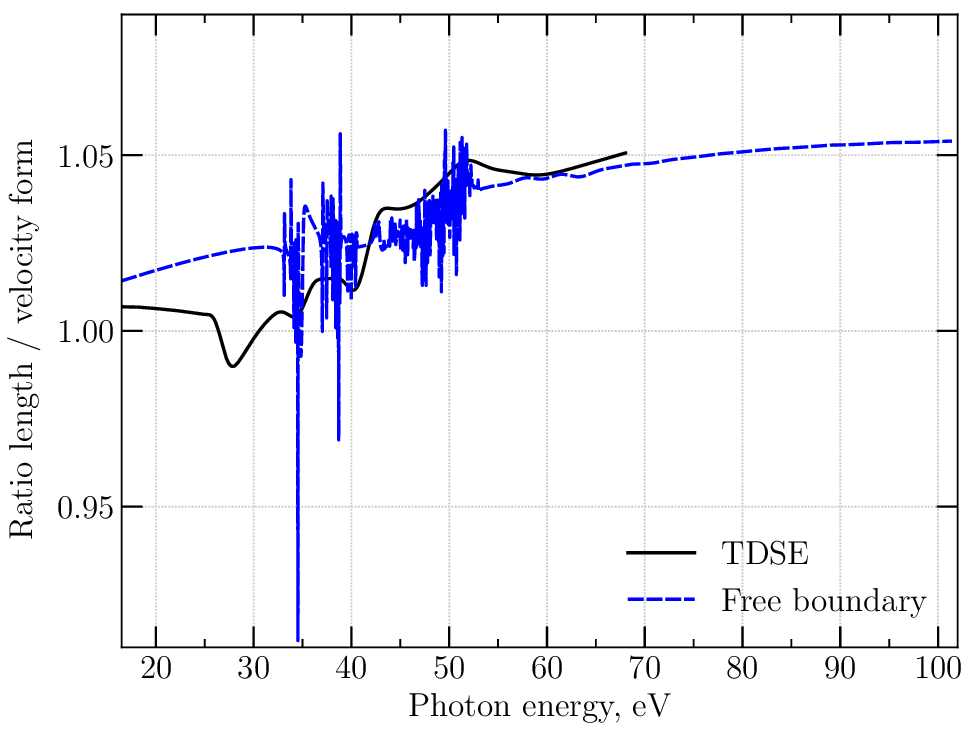}
  \caption{Ratio of the photoionization cross sections for H$_2$ molecules oriented perpendicular to the axis of polarization, computed between length and velocity form (for the FB method) and length and velocity gauge (for the TDSE method).
  The solid (black) line shows the result for the TDSE computation.
  Dashed (blue) depicts the same for the FB approach.} 
  \labfig{Pu_LvV}
\end{figure}

The behavior of the PICS is further investigated in \reffig{Su_log} for a larger energy range.
In order to properly capture the oscillatory behavior of the continuum wavefunctions at higher energies, the CI employed in this comparison features ionic orbitals from a larger pseudo-radial B-spline density of $N_\xi-1=300$ for a box-size of $\xi_\mrm{max} = 50$ $a_0$, together with 6 instead of 5 pseudo-angular channels per ionic channel, resulting in a total number of 9552 configurations for $^1\Sigma_\mrm{u}^+$ as well as 11,616 for $^1\Pi_\mrm{u}$, respectively.
For reference, the more recent results of Foj\'on \emph{et al.}~are also given and found to be in excellent agreement with the FB results (both of which use 5 ionic channels), corroborating their results from Ref.~\cite{sct:fojo04} where a slight disagreement with previous results of Semenov and Cherepkov~\cite{sct:seme03} had been noted.
As before, the TDSE PICS stay in close proximity to the FB results and show only slight deviations at about 90 eV.
At this energy, the $\ell=3$ elastic channel becomes dominant up until about 190 eV, where the $\ell=1$ channel begins to dominate again.
If this $\ell=3$ contribution is overestimated in a computation, a generally larger cross section as well as the slightly earlier crossing point of the channels indicated by the slope change can be explained.
Due to the comparably larger numerical effort in pursuing TDSE computations, the energy range is limited to about 130 eV in this work, which, however, is not a limitation by any \emph{a priori} circumstance.
Similar conclusions can be drawn from the CSG result.
While aligning mostly in close agreement with the other results, they only begin to show a more pronounced deviation starting at around 110 eV. This is expected, as both bases used for the CSG result are optimised for energies close to the ionization threshold.
 
As the PICS can be obtained from both the velocity and the mathematically equivalent length forms or gauges, results obtained in either way are compared in \reffig{Pu_LvV}.
The figure depicts the ratio between the PICS in length and velocity \emph{form} from the FB method and that between the length and velocity \emph{gauge} from the TDSE method.
For the FB method, the ratio mostly stays below 5\% for the energy range shown, which is at the level of convergence chosen for these computations.
The known preference of the velocity over the length form~\cite{sct:sanc97,sct:toff16} in terms of computational resources should only start to become visible at much higher energies than the ones considered here, where no clear indication for the superiority of one over the other has been found.
For the TDSE computations, however, preference of the velocity over the length gauge appears to be much stronger than for the FB method.
While the TDSE results below 70 eV for length and velocity gauges are found to mutually agree to an extent which is comparable to that seen for the FB method using the respective forms, obtaining convergence to a meaningful degree in the length gauge at energies beyond 70 eV has shown to be challenging, yielding more than 10\% relative deviation at 100 eV and beyond.

%Yet, there is a sudden bump in the length-gauge PICS beginning around 82 eV not seen in any other computation, giving rise to a discrepancy between gauges as high as around 25\%.
%This anomaly has been observed to be sensitive to the number of doubly excited configurations featured in the CI and tends to move to higher energies if less of these configurations are included.
%Interestingly, this surprising behavior only appears for TDSE computations employing the length gauge, but not in the case of the velocity gauge.

Finally, a comparison of the PICS for randomly-oriented molecules, for which experimental data are directly available, is presented in \reffig{pi_tot}.
It can be seen that, to a large extent, the FB as well as the TDSE results are in good qualitative, and, for lower energies, even quantitative agreement to the experimental data of Samson and Haddad~\cite{sct:sams94}.
The discrepancies of the FB and TDSE results are particularly large near the resonances, as no vibrational degrees of freedom (that are known~\cite{sct:sanc97a} to broaden the resonant features in the PICS if properly taken into account) have been considered. 
Interestingly, the relative deviations of both the FB results and that of Ref.~\cite{sct:fojo04} to the experimental data share the same overall behavior beyond 50 eV.
As also highlighted in Ref.~\cite{sct:fojo04}, the high-energy data points provided by the authors in Ref.~\cite{sct:sams94} are not direct experimental data themselves but extrapolations of a fit function with respect to the experimentally obtained values at energies below 50 eV.
Thus, in light of the agreement between the FB method results and that of Foj\'on \emph{et al.}~\cite{sct:fojo04}, the authors are inclined to believe that the deviations seen are indicative for an inaccuracy in the high-energy extrapolation of the experimental data rather than, \emph{e.g.}, a consequence of choosing an inadequate model.
Unfortunately, to the best knowledge of the authors, no experimental values of PICS for H$_2$ at these energies exist so far in the literature.
Therefore, no final conclusion can be drawn from the observations given.
However, considering the discussion concerning photoionization of helium atoms~\cite{sfa:star11}, it has been found that the discrepancies seen between different theoretical descriptions involving various levels of approximation are smaller than those between theory and experiment, particularly for higher photon energies.
As such, the experimental error bars given by Samson and Haddad have been conjectured to be chosen too optimistic for the case of helium and, thus, possibly also for H$_2$, indicating that more, sufficiently accurate measurements are needed in order to reconcile observations with theory.

\section{Discussion}
\labsec{discussion}

As mentioned earlier, both the TDSE as well as the FB method employed in this work share the same CI description and, therefrom, systematic limitations in terms of describing electronic correlation.
However, the evident differences in their respective formulations also result in different requirements for their choice of bases.
In particular, the energy resolution $\Delta E$ in the PICS is proportional to the simulation box size $R_\mrm{max}$ for the case of a time-dependent description while it is constant with respect to $R_\mrm{max}$ for the time-independent method, rendering the latter significantly more efficient for this kind of task.
This scaling originates from the fact that $\Delta E$ is proportional to the spectral pulse width, leading to longer propagation times during which the quantum-mechanical dispersion is able to broaden the extents of the electron wavepacket.
In order to assure that artifacts stemming from wavepacket reflections at the edge of the simulation-box are negligible, $R_\mrm{max}$ must be chosen sufficiently large, increasing the basis set and the CI series accordingly.
In turn, no such inter-dependency between $\Delta E$ and $R_\mrm{max}$ exists for the time-independent FB approach, allowing to utilize computational resources for higher resolution at the same cost.

Besides these limitations, an argument in favor of the time-dependent approach is its particular usefulness for studying the overlapping region between photo-excitation and ionization processes, including a proper treatment of nuclear motion.
In contrast to the time-independent methods, where spectra stemming from inevitably different time-independent descriptions have to be concatenated manually, this conveniently takes place in a single unified manner.
Likewise, in contrast to the FB method, no detailed channel-resolved analysis needs to be done, which is considerable for codes for which a time-propagation routine exists but no detailed scattering analysis tools are available and/or required.

\begin{figure}[t]
  \includegraphics[width=1.0\columnwidth]{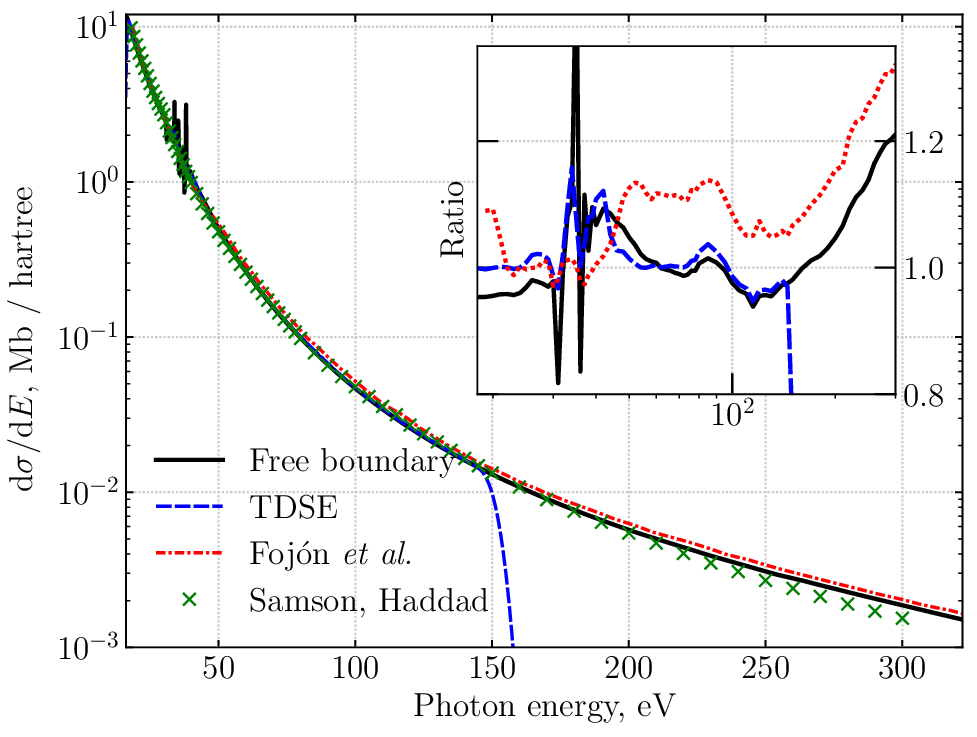}
  \caption{Total photoionization cross sections for randomly-oriented H$_2$ molecules.
  Solid (black) lines: FB-CI; dashed (blue): TDSE; dash-dotted (red): reproduced results from Foj{\'o}n \emph{et al}.~\cite{sct:fojo04}; crosses (green): experimental results from Samson and Haddad~\cite{sct:sams94}.
  The inset illustrates the ratio of all numerical results to the experimental
  reference on a logarithmic energy scale.}
  \labfig{pi_tot} 
\end{figure}

While results for the CSG methods are only presented for the $^1\Sigma_\mrm{u}^+$ symmetry in this work, there is no fundamental hindrance in implementing other symmetries as well.
Corresponding work on extending the capabilities of this approach to $^1\Pi_\mrm{u}$ symmetry is currently under way.
Judging from the $^1\Sigma_\mrm{u}^+$ results, a good agreement with the FB-CI cross sections can be expected to be found in the other cases as well. As such, the CSG method is particularly well-suited for applications where total (in contrast to partial or channel-resolved) scattering information is required.
Furthermore, the geminals allow for a very efficient inclusion of correlation and thus very accurate results.

However, as the presented comparison in \refsec{pics} has shown, the FB method turned out to be the computationally most flexible, owing to the full control offered over the ionic basis set and the supported CI series.
This allows for performing computations for large ranges of energy and processes where double-ionization channels are of minor overall importance.
In particular, it is also conceivable to extract angle-resolved information such as doubly-differential cross sections apart from the single-differential data presented in this work.
This comes at the expense of an increased complexity in properly dissecting the different coupled-channel contributions and in choosing the relevant CI series for the process at hand.
Extensions of the method are currently under active development with regards to energy-loss by electron-impact ionization and $\upbeta$-decay final-state distribution of diatomic tritium, directly
relevant for the analysis of the KATRIN experiment.

%\newpage
\section{Summary}\labsec{summary}

In this work, a comparative study of computational methods for computing single-photon ionization processes of molecular hydrogen has been given.
The obtained dipole sum rules and photoionization cross sections using the FB-CI, TDSE, and CSG methods have been found in good agreement to one another and, wherever possible, to theoretical and experimental literature values.
The study further corroborated existing PICS, with an increasingly better agreement to the more recent results.
All in all, it provided an overview of possible pathways to numerical modelling of photoionization dynamics, highlighting the respective capabilities and limitations of each method presented.

\subsection*{Acknowledgements}
J.S. acknowledges financial support by the German Federal Ministry of Research, Technology and Space (BMFTR) within ErUM-Pro 05A26PM3.

\appendix
\section{Configuration series}\labsec{ci_gs} 

\begin{table}
\centering
\caption{Two-electron CI configuration series used for the ground-states of the FB computations. The numbers indicate the index of the ionic single-particle orbitals ordered w.\,r.\,t.~increasing energy including both bound and box-discretized continuum states.}
\begin{tabular}{llll}
  \toprule 
  $e^-_1$ basis & \hspace{1.4cm}$\otimes$\hspace{0.7cm} & $e^-_2$ basis & \\
  \midrule 
  series & sym. & series & sym. \\
  \midrule 
   1 --  60 & s$\sigma_g$ & 1  --  150 & s$\sigma_g$ \\
   1 --  30 & s$\sigma_g$ & 1  --  100 & d$\sigma_g$ \\
   1 --  30 & d$\sigma_g$ & 1  --  100 & d$\sigma_g$ \\
   1 --  40 & p$\sigma_u$ & 1 --  150 & p$\sigma_u$ \\
   \midrule
   1 --  30 & d$\pi_g$ & 1 --  110 & d$\pi_g$ \\
   1 --  60 & p$\pi_u$ & 1 --  150 & p$\pi_u$ \\
   1 --  40 & f$\pi_u$ & 1 -- 100 & f$\pi_u$ \\
   \midrule
   1 --  30 & d$\delta_g$ & 1 -- 110 & d$\delta_g$ \\
   \bottomrule
\end{tabular}
\labtbl{gsci}
\end{table}

\begin{table}[b]
\caption{Two-electron CI configuration series used for the $^1\Sigma^+_\mrm{u}$ states of the TDSE computations. The numbers indicate the index of the ionic single-particle orbitals ordered w.\,r.\,t.~increasing energy including both bound and box-discretized continuum states.}
\begin{tabular}{llll}
  \toprule 
  e${}^-_1$ basis & \hspace{1.6cm}$\otimes$\hspace{0.4cm} & e${}^-_2$ basis& \\
  \midrule 
  series & sym. & series & sym. \\
  \midrule 
  % sg + su     
  1 -- \phantom{00}3 & s$\sigma_g$ & 1 -- 300 & (s,d,g,i,k)$\sigma_u$ \\
  %1 -   3 & s$\sigma_g$ & $\otimes$ & 1 - 300 & s$\sigma_u$ \\
  %1 -   3 & s$\sigma_g$ & $\otimes$ & 1 - 300 & d$\sigma_u$ \\
  %1 -   3 & s$\sigma_g$ & $\otimes$ & 1 - 300 & g$\sigma_u$ \\
  %1 -   3 & s$\sigma_g$ & $\otimes$ & 1 - 300 & i$\sigma_u$ \\
  %1 -   3 & s$\sigma_g$ & $\otimes$ & 1 - 300 & k$\sigma_u$ \\
  1 -- 300 & (s,d,g,i,k)$\sigma_g$ &  1 --  \phantom{11}3 & s$\sigma_u$ \\
  %1 - 300 & s$\sigma_g$ & $\otimes$ & 1 -   3 & s$\sigma_u$ \\
  %1 - 300 & d$\sigma_g$ & $\otimes$ & 1 -   3 & s$\sigma_u$ \\
  %1 - 300 & g$\sigma_g$ & $\otimes$ & 1 -   3 & s$\sigma_u$ \\
  %1 - 300 & i$\sigma_g$ & $\otimes$ & 1 -   3 & s$\sigma_u$ \\
  %1 - 300 & k$\sigma_g$ & $\otimes$ & 1 -   3 & s$\sigma_u$ \\
  1 -- \phantom{1}12 & s$\sigma_g$ & 1 -- \phantom{1}12 & s$\sigma_u$ \\
  1 -- \phantom{1}12 & d$\sigma_g$ & 1 -- \phantom{1}12 & d$\sigma_u$ \\
  1 -- \phantom{1}12 & s$\sigma_g$ & 1 -- \phantom{1}12 & d$\sigma_u$ \\
  1 -- \phantom{1}12 & d$\sigma_g$ & 1 -- \phantom{1}12 & s$\sigma_u$ \\
  \midrule
  % pg + pu     
  1 --  300 & (s,d,g,i,k)$\pi_g$   & 1  -- \phantom{10}2 & p$\pi_u$ \\
  %1 -  300 & s$\pi_g$   & $\otimes$ & 1  -  2  & p$\pi_u$ \\
  %1 -  300 & d$\pi_g$   & $\otimes$ & 1  -  2  & p$\pi_u$ \\
  %1 -  300 & g$\pi_g$   & $\otimes$ & 1  -  2  & p$\pi_u$ \\
  %1 -  300 & i$\pi_g$   & $\otimes$ & 1  -  2  & p$\pi_u$ \\
  %1 -  300 & k$\pi_g$   & $\otimes$ & 1  -  2  & p$\pi_u$ \\
  1 --  \phantom{1}12 & d$\pi_g$   & 1  -- \phantom{1}12  & p$\pi_u$ \\
  1 --  \phantom{1}12 & g$\pi_g$   & 1  -- \phantom{1}12  & f$\pi_u$ \\
  1 --  \phantom{1}12 & d$\pi_g$   & 1  -- \phantom{1}12  & f$\pi_u$ \\
  1 --  \phantom{1}12 & g$\pi_g$   & 1  -- \phantom{1}12  & d$\pi_u$ \\
   \bottomrule
\end{tabular}
\labtbl{tdse_ci_spu}
\end{table}

The ionic-orbital configuration series of the ground-state wavefunctions provided for the FB-PICS computations are given for each molecular symmetry in \reftbl{gsci}.
The choice of the adopted symmetries has been determined heuristically by analyzing the largest CI populations of the ground state obtained from a \emph{full}-CI computation using a significantly smaller box size.
The choice of the series truncation is loosely resembling the same strategy, but has not been particularly optimized further, as it turned out not to be the limiting factor in the overall convergence of the PICS.

Taking over-counting due to the symmetrization in \refeq{ci} into account, this results in a total number of 34,195 configurations.
Note, in contrast to the ionic basis featuring 200 orbitals for a box size of $\xi_\mrm{max}=100$ $a_0$ employed in \reffig{Su_v_Pu}, the single-particle basis used for the high-energy computation shown in Figs.~(\ref{fig:Su_log}-\ref{fig:pi_tot}) features 300 ionic orbitals per symmetry for a box size of $\xi_\mrm{max} = 50$ $a_0$, hence the truncated configuration series given in \reftbl{gsci} are further away from the full-CI limit for the smaller than for the larger box calculation.
Still, due to the smaller box size, a larger proportion of discretized continuum states is included in the CI series for the high-energy single-particle basis, leading to an overall improved description of the decaying tail behavior of the ground-state wavefunction with energy of around $-1.1734$ a.u., to be compared to the value of $-1.1722$ a.u.~obtained with the box size $\xi_\mrm{max}=100$ $a_0$.

The ionic-orbital configuration series of the $^1\Sigma^+_\mrm{u}$ symmetry is given in \reftbl{tdse_ci_spu}. The chosen elements stem from the main channels contributing to the $^1\Sigma^+_\mrm{u}$ symmetry.
Furthermore, the low-lying doubly-excited configurations are included in order to improve the description of the autoionizing states.
Note, compared to the ground-state CI from \reftbl{gsci}, so-called \enquote{complete series} are used.
The reason for this is rooted in the requirement for a proper continuum description, \emph{i.\,e.}, the inclusion of the complete series of box-discretized states in the basis.

\begingroup
\emergencystretch=1em
\bibliography{journals_v2,sfa,bsp,dia,sct,nu,csm,sfm}

@article{bsp:bros92a,
    AUTHOR  = "M. Brosolo and P. Decleva and A. Lisini",
    TITLE   = "Continuum wavefunctions calculations with
               least-squares schemes in a {B}-splines basis",
    DOI     = "10.1016/0010-4655(92)90009-N",
    JOURNAL = CPC,
    VOLUME  = 71,
    PAGES   = "207",
    YEAR    = 1992
}

@article{bsp:bros92b,
    AUTHOR  = "M. Brosolo and P. Decleva",
    TITLE   = "Variational approach to continuum orbitals in a spline
               basis: {A}n application to {H$_2^+$} photoionization",
    DOI     = "10.1016/0301-0104(92)80069-8",
    JOURNAL = CP,
    VOLUME  = 159,
    PAGES   = "185",
    YEAR    = 1992
}

@article{bsp:mart99,
    AUTHOR  = "F. Mart{\'i}n",
    TITLE   = "Ionization and dissociation using {B} splines: 
               photoionization of the hydrogen molecule",
    DOI     = "10.1088/0953-4075/32/16/201",
    JOURNAL = JPB,
    VOLUME  = 32,
    PAGES   = "R197",
    YEAR    = 1999
}

@article{bsp:bach01,
    AUTHOR  = "H. Bachau and E Cormier and P Decleva and J. E. Hansen 
               and F Mart{\'i}n",
    TITLE   = "Applications of {B}-splines in atomic and molecular physics",
    DOI     = "10.1088/0034-4885/64/12/205",
    JOURNAL = RPP,
    VOLUME  = 64,
    PAGES   = 1815,
    YEAR    = 2001
}

@article{bsp:hart02,
  title = {Regularities and irregularities in partial photoionization cross sections of {H}e},
  author = {van der Hart, H.~W. and Greene, C.~H.},
  doi = {10.1103/PhysRevA.66.022710},
  journal = PRA,
  volume = {66},
  pages = {022710},
  year = {2002},
}

@article{bsp:vann04,
    AUTHOR  = "Yulian V.~Vanne and Alejandro Saenz",
    TITLE   = "Numerical treatment of diatomic two-electron molecules using
               a {B}-spline based {CI} method",
    DOI     = "10.1088/0953-4075/37/20/005",
    JOURNAL = JPB,
    VOLUME  = 37,
    PAGES   = 4101,
    YEAR    = 2004
}

@article{bsp:dumi07,
    AUTHOR  = {I. Dumitriu and Y. V. Vanne and M. Awasthi and A. Saenz},
    TITLE   = {Photoionization of the alkali dimer cations {Li$_2{}^+$},  {Na$_2{}^+$}, and {LiNa$^+$}},
    DOI     = "10.1088/0953-4075/40/10/016",
    JOURNAL = JPB,
    VOLUME  = 40,
    PAGES   = 1821,
    YEAR    = 2007
}

@article{csm:agui71,
  author = {Aguilar, J. and Combes, J.},
  doi     = {10.1007/bf01877510},
  journal = cmp,
  pages = {269},
  title = {{A} class of analytic perturbations for one-body {S}chr{\"o}dinger {H}amiltonians},
  volume = {22},
  year = {1971}
}

@article{csm:bals71,
  author = {Balslev, E. and Combes, J.},
  doi     = {10.1007/bf01877511},
  journal = cmp,
  pages = {280},
  title = {{Spectral properties of many-body {S}chr{\"o}dinger operators with dilatation-analytic interactions}},
  volume = {22},
  year = {1971}
}

@article{csm:froe84a,
  author = {Froelich, P. and Weyrich, W.},
  doi     = {10.1063/1.446634},
  journal = jcp,
  pages = {5669},
  title = {{N}onrelativistic {C}ompton scattering in {F}urry\textquotesingle s picture: {B}eyond the sudden impulse approximation by means of the complex coordinate method},
  volume = {80},
  year = {1984}
}

@article{csm:nutt69,
  author = {Nuttall, J. and Cohen, H.~L.},
  doi = {10.1103/PhysRev.188.1542},
  issue = {4},
  journal = prl,
  month = {Dec},
  pages = {1542--1544},
  publisher = {American Physical Society},
  title = {{M}ethod of {C}omplex {C}oordinates for {T}hree-{B}ody {C}alculations above the {B}reakup {T}hreshold},
  volume = {188},
  year = {1969}
}

@article{csm:rein82,
  author = {William Reinhardt, P.},
  doi     = {10.1146/annurev.pc.33.100182.001255},
  journal = arpc,
  pages = {223},
  title = {{C}omplex {C}oordinates in the {T}heory of {A}tomic and {M}olecular {S}tructure and {D}ynamics},
  volume = {33},
  year = {1982}
}

@article{csm:resc75,
  author = {Thomas Rescigno, N. and McKoy, V.},
  doi     = {10.1103/physreva.12.522},
  journal = pra,
  pages = {522},
  title = {{R}igorous method for computing photoabsorption cross sections from a basis-set expansion},
  volume = {12},
  year = {1975}
}

@article{dia:kolo64,
 author = {W. Ko{\l}os and L. Wolniewicz},
 journal = JCP,
 doi = {10.1063/1.1725797},
 pages = {3674},
 title = {{Accurate Computation of Vibronic Energies and of Some Expectation Values for {H$_2$, D$_2$, and T$_2$}}},
 volume = {41},
 year = {1964}
}

@article{dia:kolo65,
 author = {W. Ko{\l}os and L. Wolniewicz},
 journal = JCP,
 doi = {10.1063/1.1697142},
 pages = {2429},
 title = {{Potential energy curves for the {X\,$^1\Sigma_g^+$}, 
{b\,$^3\Sigma_u^+$}, and {C\,$^1\Pi_u^+$} states of the hydrogen 
molecule}},
 volume = {43},
 year = {1965}
}

@article{dia:kolo68,
 author = {W. Ko{\l}os and L. Wolniewicz},
 journal = jcomp,
 doi = {10.1063/1.1669836},
 number = {1},
 pages = {404-410},
 title = {{Improved Theoretical Ground-State Energy of the Hydrogen Molecule}},
 volume = {49},
 year = {1968}
}

@article{dia:saen03,
 author = {A. Saenz},
 journal = PRA,
 doi = {10.1103/PhysRevA.67.033409},
 pages = {033409},
 title = {{Photoabsorption and Photoionization of {HeH}$^+$}},
 volume = {67},
 year = {2003}
}

@article{dia:pach09,
 author = {K. Pachucki},
 journal = PRA,
 doi = {10.1103/PhysRevA.80.032520},
 number = {032520},
 title = {{Two-center two-electron integrals with exponential functions}},
 volume = {80},
 year = {2009}
}

@article{dia:pach16,
 author = {K. Pachucki and M. Zientkiewizc and V. Yerokhin},
 journal = cpc,
 doi = {10.1016/j.cpc.2016.07.024},
 pages = {162-168},
 title = {{$\mathrm{H2SOLV:}$ Fortran solver for diatomic molecules in explicitly correlated exponential basis}},
 volume = {208},
 year = {2016}
}

@article{dia:brum26,
  title = {Excited $\ensuremath{\Sigma}$ states of the hydrogen-antihydrogen molecule},
  author = {Brumm, L. and Schürmann, J. and Saenz, A.},
  journal = pra,
  pages = {062801},
  volume = {113},
  year = {2026},
  month = {May},
  publisher = {American Physical Society},
  doi = {10.1103/x3xl-sxzr},
  url = {https://link.aps.org/doi/10.1103/x3xl-sxzr}
}

@string{ADNDT="Atomic Data and Nuclear Data Tables"}

@string{ADP="Ann.\,der\,Phys."}

@string{APJ="Astrophys.\,J."}

@string{ARPC="Annu.\,Rev.\,Phys.\,Chem."}

@string{CMP="Comm.\,Math.\,Phys."}

@string{CP="Chem.\,Phys."}

@string{CPC="Comp.\,Phys.\,Comm."}

@string{CPL="Chem.\,Phys.\,Lett."}

@string{EPJC="Eur.\,Phys.\,J.\,C"}

@string{FBS="Few-Body\,Syst."}

@string{IJQC="Int.\,J.\,Quant.\,Chem."}

@string{JCOMP="J.\,Comp.\,Phys."}

@string{JCP="J.\,Chem.\,Phys."}

@string{JCTC="J.\,Chem.\,Theory Comput."}

@string{JESRP="J.\,Electr.\,Spectros.\,Relat.\,Phenom."}

@string{JMS="J.\,Mol.\,Spectrosc."}

@string{JOSAB="J.\,Opt.\,Soc.\,Am.\,B"}

@string{JPB="J.\,Phys.\,B"}

@string{JPCRD="J.\,Phys.\,Chem.\,Ref.\,Data"}

@string{NC="Nat.\,Commun."}

@string{PLA="Phys.\,Lett.\,A"}

@string{PPS="Proc.\,Phys.\,Soc."}

@string{PR="Phys.\,Rev."}

@string{PRA="Phys.\,Rev.\,A"}

@string{PRE="Phys.\,Rev.\,E"}

@string{PRL="Phys.\,Rev.\,Lett."}

@string{PRP="Phys.\,Rep."}

@string{RPC="Rad.\,Phys.\,Chem."}

@string{RPP="Rep.\,Prog.\,Phys."}

@string{Science="Science"}

@string{ZP="Z.\,Phys."}

@Article{nu:froe93,
  Author                   = {P. Froelich and B. Jeziorski and W. Ko{\l}os and 
 H. Monkhorst and A. Saenz and K. Szalewicz},
  Journal                  = PRL,
  Doi                      = {10.1103/PhysRevLett.71.2871},
  Title                    = {Probability distribution of excitations to the electronic continuum of {HeT}$^+$ following the $\upbeta$ decay of the {T}$_2$ molecule},
  Year                     = {1993},
  Pages                    = {2871},
  Volume                   = {71}
}

@article{nu:katr19,
  title = {{I}mproved {U}pper {L}imit on the {N}eutrino {M}ass from a {D}irect {K}inematic {M}ethod by {KATRIN}},
  author = {Aker, M. and others},
  collaboration = {The KATRIN collaboration},
  doi = {10.1103/PhysRevLett.123.221802},
  journal = prl,
  volume = {123},
  pages = {221802},
  year = {2019},
}

@article{nu:katr25,
author = {Aker, M. and others},
collaboration = {The KATRIN collaboration},
title = {{D}irect neutrino-mass measurement based on 259 days of {KATRIN} data},
doi = {10.1126/science.adq9592},
journal = {Science},
volume = {388},
number = {6743},
pages = {180-185},
year = {2025},
}

@article{sct:aker21,
  author = {Aker, M. and others},
  collaboration = {The KATRIN Collaboration},
  doi     = {10.1063/1.1457686},
  journal = epjc,
  pages = {579},
  title = {{P}recision measurement of the electron energy-loss function in tritium and deuterium gas for the {KATRIN} experiment},
  volume = {81},
  year = {2021}
}

@article{sct:apal02,
  author = {Apalategui, A. and Saenz, A.},
  doi = {10.1088/0953-4075/35/8/309},
  journal = jpb,
  month = {apr},
  pages = {1909},
  publisher = {},
  title = {Multiphoton ionization of the hydrogen molecule {H}$_2$},
  volume = {35},
  year = {2002}
}

@article{sct:awas05,
  author = {Awasthi, M. and Vanne, Y.~V. and Saenz, A.},
  doi = {10.1088/0953-4075/38/22/005},
  journal = jpb,
  month = {oct},
  pages = {3973},
  publisher = {},
  title = {{Non-perturbative solution of the time-dependent {S}chrödinger equation describing {H}$_2$ in intense short laser pulses}},
  volume = {38},
  year = {2005}
}

@article{sct:back76,
  author = {Backx, C. and Wight, G.~R. and der Wiel, M.~J.~V.},
  doi     = {10.1088/0022-3700/9/2/018},
  journal = jpb,
  pages = {315},
  title = {{O}scillator strengths (10--70 e{V}) for absorption, ionization and dissociation in {H}$_2$, {HD} and {D}$_2$, obtained by an electron-ion coincidence method},
  volume = {9},
  year = {1976}
}

@article{sct:beth30,
  author = {Bethe, H.},
  doi     = {10.1002/andp.19303970303},
  journal = adp,
  pages = {325},
  title = {{Z}ur {T}heorie des {D}urchgangs schneller {K}orpuskularstrahlen durch {M}aterie},
  volume = {397},
  year = {1930}
}

@article{sct:borb07,
  author = {Borb{\'e}ly, S. and Nagy, L.},
  doi     = {10.1016/j.radphyschem.2006.01.035},
  journal = rpc,
  pages = {516--520},
  title = {{S}tudy of the interference effects in the ionization of {H}$_2$ by the use of two-center wavefunctions},
  volume = {76},
  year = {2007}
}

@article{sct:brag92,
  author = {Brage, T. and Fischer, C.~F. and Miecznik, G.},
  doi     = {10.1088/0953-4075/25/24/010},
  journal = jpb,
  pages = {5289},
  title = {{N}on-variational, spline-{G}alerkin calculations of resonance positions and widths, and photodetachment and photo-ionization cross sections for {H}$^-$ and {H}e},
  volume = {25},
  year = {1992}
}

@article{sct:cace93,
  author = {Cacelli, I. and Moccia, R. and Rizzo, A.},
  doi     = {10.1016/s0301-0104(99)00325-0},
  journal = jcp,
  pages = {8742--8748},
  title = {{Gaussian type orbitals basis sets for the calculation
of continuum properties in molecules: 
{The} photoionization cross section of {H}$_2$}},
  volume = {98},
  year = {1993}
}

@article{sct:cohe66,
  author = {Cohen, H.~D. and Fano, U.},
  doi = {10.1103/PhysRev.150.30},
  journal = pr,
  pages = {30--33},
  title = {{I}nterference in the {P}hoto-{I}onization of {M}olecules},
  volume = {150},
  year = {1966}
}

@article{sct:coll84,
  author = {Collins, L.~A. and Schneider, B.~I.},
  doi     = {10.1088/0022-3700/14/3/013},
  journal = pra,
  pages = {1695--1708},
  title = {{M}olecular photoionization in the linear algebraic approach: {H}$_{2}$, {N}$_{2}$, {NO}, and {CO}$_{2}$},
  volume = {29},
  year = {1984}
}

@article{sct:flan65,
  author = {Flannery, M.~R. and {\"O}pik, U.},
  doi     = {10.1088/0370-1328/86/3/307},
  journal = pps,
  pages = {491},
  title = {{T}he photoionization of the hydrogen molecule from the ground electronic and vibrational state},
  volume = {86},
  year = {1965}
}

@article{sct:flan77,
  author = {Flannery, M. and Tai, H. and Albritton, D.},
  doi     = {10.1016/0092-640x(77)90039-0},
  journal = adndt,
  pages = {563--585},
  title = {{C}ross sections for the photoionization of {H}$_2$({X}${}^1{\Sigma_g}^+,\,\nu_i=0$--14) with the formation of {H}$_2{}^+$({X}${}^2{\Sigma}_g^+,\,\nu_f=0$--18), and vibrational overlaps and {R}$^n$-centroids for the associated vibrational transitions},
  volume = {20},
  year = {1977}
}

@article{sct:fojo04,
  author = {Foj\'on, O.~A. and Fern\'andez, J. and Palacios, A. and Rivarola, R.~D. and Mart\'in, F.},
  doi     = {10.1088/0953-4075/37/15/003},
  journal = jpb,
  pages = {3035},
  title = {Interference effects in {H}$_2$ photoionization at high energies},
  volume = {37},
  year = {2004}
}

@article{sct:ford75,
  author = {Ford, A.~L. and Docken, K.~K. and Dalgarno, A.},
  doi     = {10.1086/153387},
  journal = apj,
  pages = {819},
  title = {{T}he photoionization and dissociative photoionization of {H}$_2$, {HD}, and {D}$_2$},
  volume = {195},
  year = {1975}
}

@article{sct:gall88,
  author = {Gallagher, J.~W. and Brion, C.~E. and Samson, J.~A.~R. and Langhoff, P.~W.},
  doi     = {10.1063/1.555821},
  journal = jpcrd,
  pages = {9--153},
  title = {{A}bsolute {C}ross {S}ections for {M}olecular {P}hotoabsorption, {P}artial {P}hotoionization, and {I}onic {P}hotofragmentation {P}rocesses},
  volume = {17},
  year = {1988}
}

@article{sct:glas07,
  author = {Glass-Maujean, M. and Klumpp, S. and Werner, L. and Ehresmann, A. and Schmoranzer, H.},
  doi     = {10.1063/1.2435345},
  journal = jcp,
  pages = {094306},
  title = {{C}ross sections for the ionization continuum of {H}$_2$ in the 15.3--17.2 e{V} energy range},
  volume = {126},
  year = {2007}
}

@article{sct:gord28,
  author = {Gordon, W.},
  doi     = {10.1007/bf01351302},
  journal = zp,
  pages = {180--191},
  title = {{\"U}ber den {S}to{\ss} zweier {P}unktladungen nach der {W}ellenmechanik},
  volume = {48},
  year = {1928}
}

@article{sct:hara86,
  author = {Hara, S. and Sato, H. and Ogata, S. and Tamba, N.},
  doi     = {10.1088/0022-3700/19/8/013},
  journal = jpb,
  pages = {1177},
  title = {{V}ibrationally and rotationally resolved cross sections and angular distributions of photoelectrons from {H}$_2$},
  volume = {19},
  year = {1986}
}

@article{sct:itik83,
  author = {Itikawa, Y. and Takagi, H. and Nakamura, H. and Sato, H.},
  doi     = {10.1103/physreva.27.1319},
  journal = pra,
  pages = {1319--1327},
  title = {{T}heoretical studies of photoionization of hydrogen molecules},
  volume = {27},
  year = {1983}
}

@article{sct:khar68,
  author = {Khare, S.~P.},
  doi     = {10.1103/physrev.173.43},
  journal = pre,
  pages = {43--49},
  title = {{P}hoto-{I}onization of the {H}ydrogen {M}olecule},
  volume = {173},
  year = {1968}
}

@article{sct:koss89,
  author = {Kossmann, H. and Schwarzkopf, O. and Kammerling, B. and Braun, W. and Schmidt, V.},
  doi     = {10.1088/0953-4075/22/14/004},
  journal = jpb,
  pages = {L411},
  title = {{P}hotoionisation cross section of {H}$_2$},
  volume = {22},
  year = {1989}
}

@article{sct:lamb98,
  author = {Lambropoulos, P. and Maragakis, P. and Zhang, J.},
  doi     = {10.1016/s0370-1573(98)00027-1},
  journal = prp,
  pages = {203--293},
  title = {{T}wo-electron atoms in strong fields},
  volume = {305},
  year = {1998}
}

@article{sct:lati95,
  author = {Latimer, C.~J. and Dunn, K.~F. and O’Neill, F.~P. and MacDonald, M.~A. and Kouchi, N.},
  doi     = {10.1063/1.469185},
  journal = jcp,
  pages = {722--725},
  title = {{P}hotoionization of hydrogen and deuterium},
  volume = {102},
  year = {1995}
}

@article{sct:liu04,
  author = {Liu, X. and Donald Shemansky, E.},
  doi = {10.1086/423890},
  journal = {Astophys.~J.},
  month = {oct},
  pages = {1132},
  title = {{I}onization of {M}olecular {H}ydrogen},
  volume = {614},
  year = {2004}
}

@article{sct:lucc81,
  author = {Lucchese, R.~R. and McKoy, V.},
  doi     = {10.1103/physreva.24.770},
  journal = pra,
  pages = {770--776},
  title = {{I}terative approach to the {S}chwinger variational principle applied to electron---molecular-ion collisions},
  volume = {24},
  year = {1981}
}

@article{sct:mart74,
  author = {Martin, P. and Rescigno, T. and McKoy, V. and Henneker, W.},
  doi     = {10.1016/0009-2614(74)85077-3},
  journal = cpl,
  pages = {496--501},
  title = {{P}hotoionization cross sections for {H}$_2$ in the random phase approximation with a square-integrable basis},
  volume = {29},
  year = {1974}
}

@article{sct:nagy04,
  author = {Nagy, L. and Borb{\'e}ly, S. and P{\'o}ra, K.},
  doi     = {10.1016/j.physleta.2004.06.001},
  journal = pla,
  pages = {481--489},
  title = {{I}nterference effects in the photoionization of molecular hydrogen},
  volume = {327},
  year = {2004}
}

@article{sct:niko06,
  author = {Nikolopoulos, L.~A.~A.},
  doi     = {10.1103/physreva.73.043408},
  journal = pra,
  pages = {043408},
  title = {{E}lectromagnetic transitions between states satisfying free-boundary conditions},
  volume = {73},
  year = {2006}
}

@article{sct:onei78,
  author = {ONeil, S.~V. and Reinhardt, W.~P.},
  doi     = {10.1063/1.436813},
  journal = jcp,
  pages = {2126--2142},
  title = {{P}hotoionization of molecular hydrogen},
  volume = {69},
  year = {1978}
}

@article{sct:rase83,
  author = {Ra\c{s}eev, G. and Le Rouzo, H.},
  doi     = {10.1103/physreva.27.268},
  journal = pra,
  pages = {268--284},
  title = {{E}lectronic ab initio quantum-defect theory. {L}ow-resolution {H}$_{2}$ photoionization spectrum},
  volume = {27},
  year = {1983}
}

@article{sct:rase84,
  author = {Ra\c{s}eev, G.},
  doi     = {10.1088/0022-3700/18/3/018},
  journal = jpb,
  pages = {423--439},
  title = {{V}ariational calculation of the logarithmic derivative of the wavefunction: the electronic autoionisation region in photoionisation of {H}$_2$.},
  volume = {18},
  year = {1984}
}

@article{sct:rein79,
  author = {Reinhardt, W.~P.},
  doi     = {10.1016/0010-4655(79)90064-x},
  journal = cpc,
  pages = {1--21},
  title = {{L}$^2$ discretization of atomic and molecular electronic continua: {M}oment, quadrature and {J}-matrix techniques},
  volume = {17},
  year = {1979}
}

@article{sct:rich84,
  author = {Richards, J.~A. and Larkins, F.~P.},
  doi     = {10.1088/0022-3700/17/6/015},
  journal = jpb,
  pages = {1015--1026},
  title = {{M}olecular photoionisation calculations with numerical continuum wavefunctions: application to the hydrogen molecule},
  volume = {17},
  year = {1984}
}

@article{sct:rich86,
  author = {Richards, J.~A. and Larkins, F.~P.},
  doi     = {10.1088/0022-3700/19/13/008},
  journal = jpb,
  pages = {1945},
  title = {{P}hotoionisation cross section calculations of $\mathrm{H}_2$ and $\mathrm{H}_2{}^+$ using numerical continuum wavefunctions},
  volume = {19},
  year = {1986}
}

@article{sct:saen93,
  author = {Saenz, A. and Weyrich, W. and Froelich, P.},
  doi = {10.1002/qua.560460304},
  journal = ijqc,
  pages = {365--374},
  title = {{A} configuration-interaction-oriented implementation of the complex coordinate method},
  volume = {46},
  year = {1993}
}

@article{sct:saen96,
  author = {Saenz, A. and Weyrich, W. and Froelich, P.},
  doi = {10.1088/0953-4075/29/1/014},
  journal = jpb,
  pages = {97--113},
  title = {{T}he first {B}orn approximation and absolute scattering cross sections},
  volume = {29},
  year = {1996}
}

@article{sct:sams94,
  author = {Samson, ~.~J. and Haddad, G.~N.},
  doi = {10.1364/JOSAB.11.000277},
  journal = josab,
  month = {Feb},
  pages = {277--279},
  publisher = {Optica Publishing Group},
  title = {Total photoabsorption cross sections of {H}$_2$ from 18 to 113 e{V}},
  volume = {11},
  year = {1994}
}

@article{sct:sanc97,
  author = {S{\'a}nchez, I. and Mart{\'i}n, F.},
  doi = {10.1088/0953-4075/30/3/021},
  journal = jpb,
  month = {feb},
  pages = {679},
  publisher = {},
  title = {Representation of the electronic continuum of {H}$_2$  with  {B}-spline basis},
  volume = {30},
  year = {1997}
}

@article{sct:sanc97a,
  author = {S{\'a}nchez, I. and Mart{\'i}n, F.},
  doi     = {10.1103/physreva.21.1480},
  journal = jcp,
  pages = {8391--8396},
  title = {Resonant effects in photoionization of {H}$_2$ and {D}$_2$},
  volume = {107},
  year = {1997}
}

@article{sct:sanz07,
  author = {Sanz-Vicario, J. and Palacios, A. and Cardona, J. and Bachau, H. and Mart{\'i}n, F.},
  doi     = {10.1016/j.elspec.2007.02.011},
  journal = jesrp,
  pages = {182--187},
  title = {{A}b initio time-dependent method to study the hydrogen molecule exposed to intense ultrashort laser pulses},
  volume = {161},
  year = {2007}
}

@article{sct:schn24,
  author = {Schneidewind, S. and Sch{\"u}rmann, J. and Lokhov, A. and Weinheimer, C. and Saenz, A.},
  doi = {10.1140/epjc/s10052-024-12802-w},
  journal = epjc,
  month = {05},
  pages = {},
  title = {Improved treatment of the {T}$_2$ molecular final-states uncertainties for the {KATRIN} neutrino-mass measurement},
  volume = {84},
  year = {2024}
}

@article{sct:seme03,
  author = {Semenov, S.~K. and Cherepkov, N.~A.},
  doi     = {10.1088/0953-4075/36/7/310},
  journal = jpb,
  pages = {1409},
  title = {Photoionization of the {H}$_2$ molecule in the random phase approximation},
  volume = {36},
  year = {2003}
}

@article{sct:stas02,
  author = {Staszewska, G. and Wolniewicz, L.},
  doi     = {10.1006/jmsp.2002.8546},
  issn = {0022-2852},
  journal = jms,
  pages = {208--212},
  title = {{A}diabatic {E}nergies of {E}xcited ${}^1{\Sigma}_u$ {S}tates of the {H}ydrogen {M}olecule},
  volume = {212},
  year = {2002}
}

@article{sct:tenn10,
  author = {Tennyson, J.},
  doi     = {10.1016/j.physrep.2010.02.001},
  journal = prp,
  pages = {29--76},
  title = {{E}lectron--molecule collision calculations using the {R}-matrix method},
  volume = {491},
  year = {2010}
}

@article{sct:tenn86,
  author = {Tennyson, J. and Noble, C.~J. and Burke, P.~G.},
  doi     = {10.1002/qua.560290502},
  journal = ijqc,
  pages = {1033--1042},
  title = {{C}ontinuum states of the hydrogen molecule with the {R}‐{M}atrix method},
  volume = {29},
  year = {1986}
}

@article{sct:toff16,
  author = {Toffoli, D. and Decleva, P.},
  doi     = {10.1021/acs.jctc.6b00627},
  journal = jctc,
  pages = {4996--5008},
  title = {{A} {M}ultichannel {L}east-{S}quares {B}-{S}pline {A}pproach to {M}olecular {P}hotoionization: {T}heory, {I}mplementation, and {A}pplications within the {C}onfiguration--{I}nteraction {S}ingles {A}pproximation},
  volume = {12},
  year = {2016}
}

@article{sct:volk26,
  author = {Volkmann, H. and Saenz, A.},
  doi     = {10.1007/s00601-026-02061-8},
  journal = fbs,
  pages = {38},
  title = {{A} {D}istorted {S}ingle-{C}enter {A}pproach to {T}wo-{C}enter {C}oulomb {S}cattering},
  volume = {67},
  year = {2026}
}

@article{sct:wait17,
  author = {Waitz, M. and Bello, R.~Y. and Metz, D. and Lower, J. and Trinter, F. and Schober, C. and Keiling, M. and Lenz, U. and Pitzer, M. and Mertens, K. and Martins, M. and Viefhaus, J. and Klumpp, S. and Weber, T. and Schmidt, L.~P.~H. and Williams, J.~B. and Sch{\"o}ffler, M.~S. and Serov, V.~V. and Kheifets, A.~S. and Argenti, L. and Palacios, A. and Mart{\'i}n, F. and Jahnke, T. and D{\"o}rner, R.},
  doi     = {10.1038/s41467-017-02437-9},
  journal = nc,
  pages = {2266},
  title = {{I}maging the square of the correlated two-electron wave function of a hydrogen molecule},
  volume = {8},
  year = {2017}
}

@article{sct:wiel72,
  author = {{van der Wiel}, M.~J. and Brion, C.~E.},
  doi     = {10.1016/0368-2048(72)80034-3},
  journal = {J. Electron Spectrosc.},
  pages = {309--318},
  title = {{\textquoteleft {P}hotoelectron\textquoteright} spectroscopy by electron impact coincidence measurements of scattered and ejected electrons in {CO}},
  url = {https://www.sciencedirect.com/science/article/pii/0368204872800343},
  volume = {1},
  year = {1972}
}

@article{sct:woln03,
  author = {Wolniewicz, L. and Staszewska, G.},
  doi     = {10.1016/s0022-2852(03)00121-8},
  issn = {0022-2852},
  journal = jms,
  pages = {45--51},
  title = {{E}xcited ${}^1{\Pi}_u$ states and the ${}^1{\Pi}_u$ $\rightarrow$ ${X}^1{\Sigma}_g^+$ transition moments of the hydrogen molecule},
  url = {https://www.sciencedirect.com/science/article/pii/S0022285203001218},
  volume = {220},
  year = {2003}
}

@article{sct:woln93,
  author = {Wolniewicz, L.},
  doi = {10.1063/1.465303},
  issn = {0021-9606},
  journal = jcp,
  month = {08},
  pages = {1851--1868},
  title = {{R}elativistic energies of the ground state of the hydrogen molecule},
  volume = {99},
  year = {1993}
}

@article{sct:woln95,
  author = {Wolniewicz, L.},
  doi = {10.1063/1.469753},
  issn = {0021-9606},
  journal = jcp,
  month = {08},
  pages = {1792--1799},
  title = {{N}onadiabatic energies of the ground state of the hydrogen molecule},
  volume = {103},
  year = {1995}
}

@article{sct:yan01,
  author = {Yan, M. and Sadeghpour, H.~R. and Dalgarno, A.},
  doi     = {10.1086/322775},
  journal = apj,
  pages = {1194},
  title = {{Erratum {P}hotoionization cross sections of {H}e and {H}$_2$}},
  volume = {559},
  year = {2001}
}

@article{sct:yan98,
  author = {Yan, M. and Sadeghpour, H.~R. and Dalgarno, A.},
  doi     = {10.1086/322775},
  journal = apj,
  pages = {1044},
  title = {{Photoionization cross sections of {H}e and {H}$_2$}},
  volume = {496},
  year = {1998}
}

@book{sct:bart96,
  address = {Berlin Heidelberg},
  author = {Bartschat, K.},
  publisher = {Springer},
  title = {{C}omputational {A}tomic {P}hysics},
  year = {1996}
}

@incollection{sct:star23,
  address = {Cham},
  author = {Starace, A.~F.},
  booktitle = {Springer Handbook of Atomic, Molecular, and Optical Physics},
  pages = {383--394},
  publisher = {Springer International Publishing},
  title = {{P}hotoionization of {A}toms},
  year = {2023}
}

@Article{sfa:niko01a,
  Title                    = {Multichannel theory of two-photon single and double
 ionization of helium},
  Author                   = {L. A. A. Nikolopoulos and P. Lambropoulos},
  Doi                      = {10.1088/0953-4075/34/4/304},
  Journal                  = JPB,
  Year                     = {2001},
  Pages                    = {545},
  Volume                   = {34}
}

@Article{sfa:star11,
  Title                    = {Accurate non-relativistic photoionization cross section
 for {He} at non-resonant photon energies},
  Author                   = {Stark, A. and Saenz, Alejandro},
  Doi                      = {10.1088/0953-4075/44/3/035004},
  Journal                  = JPB,
  Year                     = {2011},
  Pages                    = {035004},
  Volume                   = {44}
}

@Article{sfm:foer14,
  Title                    = {Ionization behavior of molecular hydrogen in intense laser fields: 
 Influence of molecular vibration and alignment},
  Author                   = {Johann F\"orster and Yulian V. Vanne and Alejandro Saenz},
  Doi                      = {10.1103/PhysRevA.90.053424},
  Journal                  = PRA,
  Year                     = {2014},
  Pages                    = {053424},
  Volume                   = {90}
}

@Article{sfm:pala07,
  Title                    = {Excitation and ionization of molecular hydrogen by ultrashort vuv 
 laser pulses},
  Author                   = {A. Palacios and H. Bachau and F. Mart{\'i}n},
  Doi                      = {10.1103/PhysRevA.75.013408},
  Journal                  = PRA,
  Year                     = {2007},
  Pages                    = {013408},
  Volume                   = {75}
}
\endgroup

\end{document}